# General relativistic hydrodynamics with a Roe solver

Frits Eulderink[1][*], Garrelt Mellema[2,1]

[1] Sterrewacht Leiden, Postbus 9513, 2300 RA Leiden, The Netherlands
[2] Astrophysics Group, Department of Mathematics, UMIST, P.O. Box 88, Manchester M60 1QD, UK
g_mellema@ast.ma.umist.ac.uk

**Abstract.** We present a numerical method to solve the equations of general relativistic hydrodynamics in a given external gravitational field. The method is based on a generalization of Roe's approximate Riemann solver for the non relativistic Euler equations in Cartesian coordinates. The new method is applied to a set of standard test problems for general relativistic hydrodynamics, and is shown to perform well in comparison to existing numerical schemes. In contrast to existing explicit methods the present method can cope with strong relativistic shocks. By-products are: the characteristic form of the general relativistic Euler equations, a numerical method for special relativity that can deal with strong discontinuities, a numerical scheme for the integration of the Euler equations in an arbitrary coordinate system, possibly under the influence of (external) gravity, and a novel method to incorporate source terms in numerical schemes.

## 1. Introduction

Astronomical jets are observed to be quite common. They consist of narrow plasma beams that transport energy from a region near a central compact object out to distances of a hundred thousand or more times the dimensions of the central object. In active galaxies the central object is likely to be a black hole and energy is transported from the inner 10pc out to 100kpc or more (Begelman et al. 1984). In the most extreme cases velocities are inferred that are in excess of 99% of the speed of light $c$. In galactic sources like SS433 the central object is likely to be a neutron star. Here energy is transported all the way to the extended radio source W50 (Vermeulen 1989), and bulk velocities of $0.26c$ are reached (Margon 1984). Hence there are both galactic and extragalactic objects in which the velocities are close enough to the speed of light that relativistic effects should be taken into account.

Strong gravitational fields may be important for jet formation and propagation close to the central object. Therefore, general relativistic effects must be considered if the central region is to be studied properly. Apart from jets there are various other astrophysical situations, including supernova collapse, where it is desirable to have solutions of the equations of general relativistic hydrodynamics. Such solutions cannot be found analytically for most initial and boundary conditions. Therefore, one has to turn to a numerical approximation to the desired solution. Some of the cases for which analytical solutions do exist are used as test problems to see how accurate a given numerical method is. An overview of relativistic hydrodynamics was written by Taub (1978).

From a mathematical point of view it is interesting to see how well the numerical concepts that were developed for the classical Euler equations apply to the special relativistic Euler equations and to the even more complex general relativistic Euler equations.

Several numerical methods for relativistic hydrodynamics exist, e.g. the Lagrangian finite difference method of Van Riper (1979), and the spectral element method of Van Putten (1992). An Eulerian finite difference method based on upwind fluxes was developed and used by Hawley, Smarr & Wilson (1984a, 1984b), henceforth HSW1 and HSW2. The main shortcoming of these methods is that they cannot deal with ultra relativistic shocks (HSW2). At some point it looked that no explicit numerical method would ever be able to deal with ultra relativistic shocks (Norman & Winkler 1986). However, major progress has been made by using schemes that use the conservative form of the equations of relativistic hydrodynamics and use 'Riemann solvers' to follow the evolution of the flow. Examples of this are to be found in Martí et al. (1991), Marquina et al. (1992), and Schneider et al. (1993). The earliest example was Eulderink (1988), which forms the basis of the present article. As we will show, this explicit method can deal with shocks in flows with speeds of up to $0.99999c$. Also, unlike the other

---

*Send offprint requests to*: G. Mellema
[*] Present address: Koninklijke/Shell-Laboratorium, Amsterdam, Postbus 3003, 1003 AA Amsterdam, the Netherlands

relativistic Riemann solvers mentioned above, we consider the general relativistic Euler equations.

Comparative tests in the non relativistic domain have shown that numerical methods that use characteristics are superior to other methods for violent flows (Woodward & Colella 1984). Experience has demonstrated that Roe's (1981b, 1986a) numerical method is particularly suited to treat strong shocks in the classical Euler equations (e.g. Chang & Liou 1988). Furthermore, this method stays close enough to the physics of the Euler equations to allow generalization to other equations that are physical generalizations of the classical Euler equations in Cartesian coordinates. The method makes no use of arbitrary parameters that must be tuned by experiment and that may not perform quite as well in the generalized situation. The artificial viscosity introduced in some methods (e.g. HSW1 and HSW2) to smooth unphysical oscillations near discontinuities is a typical example of an arbitrary parameter that is not straightforward to tune in the relativistic domain (HSW2). For these reasons Roe's method was taken as a starting point for the present method.

An important advantage of studying the general relativistic equations is that their formulation in terms of a metric is so general that two interesting limits may be taken. By inserting the Minkowski metric a new method for solving the equations of special relativistic hydrodynamics results; by taking all speeds to be small compared to the speed of light one obtains a method for the classical equations of fluid motion in an arbitrary frame of reference. This is important because three-dimensional hydrodynamics is very computer intensive by today's standards. Hence many problems can only be studied in coordinate frames in which the symmetry of the problem can be used to reduce its dimensionality. Usually this is achieved at the expense of a non-Cartesian coordinate frame. The present method can be applied in these situations.

Because it is extremely difficult to decide how close the numerical solution of a complicated flow problem is to reality, numerical methods are usually tested on easier problems to which an exact solution is known. To demonstrate the capabilities of the method presented below, it is applied to the set of test problems for general relativistic codes listed in HSW1. These tests show that the present method overcomes said problems at ultra relativistic speeds. For the one-dimensional test problems of HSW it is superior to the method of HSW at a given resolution and time step. However HSW's method is likely to be faster. Comparisons based on equal CPU time have not been made because HSW do not mention the CPU times they require, and because the CPU time depends on the machine, the implementation, and the dimensionality of the problem. Results of astrophysical applications of our method were reported by Mellema et al. (1991), Icke et al. (1992) and Eulderink & Mellema (1994).

This paper is organized as follows: in Sect. 2 the equations are introduced in the form in which they will be studied. In Sect. 3 some basic numerical concepts are introduced. Section 4 describes the Roe solver for the classical Euler equation in Cartesian coordinates. Section 5 is about ways to determine the primitive variables from the conserved state vector. Section 6 deals with the interpretation of data within a computational cell. Section 7 explains the analytical concept of characteristics in the presence of a non-constant metric. Section 8 contains the step by step recipe of the present method. Section 9 describes the function that chooses between a discontinuous and a smooth interpretation of the data in two neighbouring cells. In Sect. 10 the general relativistic Roe solver is derived. Section 11 shows limiting cases of that Roe solver, such as the special relativistic Roe solver and the classical Roe solver in non-Cartesian coordinates. In Sect. 12 the method is compared to both analytical results and to other numerical methods. Section 13 contains the discussion and conclusions.

In the rest of this paper the abbreviations C for Cartesian, NR for non relativistic, SR for special relativistic, GR for general relativistic and H for hydrodynamics are used.

## 2. Equations of relativistic hydrodynamics

The familiar equations of general relativistic fluid dynamics in the usual notation with the speed of light in vacuo equal to unity ($c = 1$), the gravitational constant times the central mass equal to unity ($GM = 1$) and the Einstein summation convention are (Misner et al. 1973)

$$(\rho u^\alpha)_{;\alpha} = 0, \qquad (2.1)$$

$$T^{\alpha\beta}_{;\beta} = S^\alpha, \qquad (2.2)$$

where ; denotes covariant differentiation, $\rho$ is the fluid density in the comoving frame, $u^\alpha$ the four-velocity of the fluid, $T^{\alpha\beta}$ the stress-energy tensor and $S^\alpha$ the source term. Since the Einstein equations are divergence free, total energy and momentum of a system are conserved. Hence just like in the NR case the source term can be expressed as the negative divergence of a tensor. The energy and momentum supplied to the fluid by radiation for example, can both be expressed as the integral of the absorbed and scattered radiation and as the negative divergence of the radiation energy-momentum tensor (Mihalas & Mihalas 1984, p. 151 and p. 430). For use in numerical hydrodynamics the latter form is usually undesirable because an extra differentiation is introduced.

In a space-like metric $g_{\alpha\beta}$ normalization of the four-velocity can be expressed by

$$g_{\alpha\beta} u^\alpha u^\beta = -1. \qquad (2.3)$$

Upon writing $g$ for the determinant of the covariant metric $g_{\alpha\beta}$ and using the identities (86.9) and (86.10) of Landau and Lifshitz (1971) one can rewrite the equations in the form that is used for numerical study below:

$$(\sqrt{-g}\rho u^\alpha)_{,\alpha} = 0 \qquad (2.4)$$

and

$$(\sqrt{-g}T^{\alpha\beta})_{,\beta} = \hat{S}^\alpha, \qquad (2.5)$$

where

$$\hat{S}^\alpha \equiv \sqrt{-g}(S^\alpha - \Gamma^\alpha_{\beta\gamma}T^{\beta\gamma}). \tag{2.6}$$

Here $\Gamma^\alpha_{\beta\gamma}$ denotes a Christoffel symbol, defined as

$$\Gamma^\alpha_{\beta\gamma} \equiv \tfrac{1}{2}g^{\alpha\delta}(g_{\delta\beta,\gamma} + g_{\gamma\delta,\beta} + g_{\beta\gamma,\delta}). \tag{2.7}$$

A more compact form of Eqs. (2.4-6) is

$$F^\alpha_{,\alpha} = \mathcal{S}, \tag{2.8}$$

in which $\mathcal{S}$ is the five-dimensional vector

$$\mathcal{S} \equiv (0, \hat{S}^\alpha)^T \equiv (0, \hat{S}^0, \hat{S}^1, \hat{S}^2, \hat{S}^3)^T. \tag{2.9}$$

Each of the four $F^\beta$ is a five-dimensional vector defined by

$$F^\beta \equiv \sqrt{-g}(\rho u^\beta, T^{\beta 0}, T^{\beta 1}, T^{\beta 2}, T^{\beta 3})^T. \tag{2.10}$$

The superscript $^T$ indicates transposition, to remind the reader that these vectors are to be used as column vectors under left multiplication by a matrix. With this understanding, we will henceforth omit this superscript.

In this form the equations are used in the remainder of this paper. Here $F^0$ is called the state vector. Each five-dimensional vector $F^i$ is a flux in the coordinate direction $x^i$. The contravariant coordinates are denoted by $x^\alpha$, and $x^0$ is loosely referred to as time. In what follows variables that are given functions of the coordinates (such as the metric) are collectively called the 'known functions'. By contrast the variables that have to be found by integrating the differential equation are called the 'unknowns'. The source vector $\mathcal{S}$ contains terms due to curvature and, possibly, "external" source terms due to e.g. radiation. Notice that there is a difference between the mathematical concept of a source term $\mathcal{S}$ and the physical concept $S^\alpha$: even if there is physically no source of energy-momentum ($S^\alpha = 0$) there may be a mathematical source term due to curvature.

Next we make specific choices for the stress-energy tensor and the source term. In most astrophysical problems one is interested in length scales that are much larger than the characteristic length scale of dissipative effects. Under those circumstances viscosity and thermal conduction may be neglected. This also holds if shocks are present, if one is interested in the effect of the shocks on the flow rather than in the precise shock structure (Landau & Lifshitz 1959 §87). Henceforth thermal conduction and viscosity are neglected in this paper. The stress-energy tensor is then that of an ideal fluid, i.e. a fluid with constant entropy along a streamline if no shocks and source terms are of influence:

$$T^{\alpha\beta} = \rho h u^\alpha u^\beta + p g^{\alpha\beta}, \tag{2.11}$$

where $p$ is the pressure in the comoving frame and $h$ the relativistic enthalpy (i.e. including rest mass energy) in the comoving frame. The entropy of the fluid may increase due to shocks or due to the source term. An example is the source term that arises in case of the interaction of a fluid with a known radiation field (Mihalas & Mihalas 1984, p. 151 and p. 430). In that case entropy is conserved only if all radiation is scattered and none is absorbed by the fluid (Lindquist 1966). In general entropy is generated unless $u_\alpha S^\alpha = 0$, as is physically obvious. In this paper the *external* source term $S^\alpha$ is introduced only for the sake of completeness. It is taken to be zero in what follows.

Equations (2.8) and (2.11) need to be complemented by an equation of state. For a polytropic gas with adiabatic exponent $\Gamma$, it is given by

$$\rho h = \rho + \frac{\Gamma}{\Gamma - 1}p \tag{2.12}$$

(e.g. Courant & Friedrichs 1967, Sec. 4).

Equation (2.8) is a hyperbolic set of partial differential equations. This is the mathematical manifestation of the fact that information takes time to spread in space. In this case this occurs through motion of the fluid itself and through sound waves (see Courant & Friedrichs 1967). An immediate consequence of a finite speed of communication is the possibility of discontinuities in the solution. In fluid dynamics, these often have a physical meaning. They are not only introduced through the initial state of a fluid but can also be created from smoother flows by nonlinearities. Of course real fluids behave smoothly. However the scale over which viscosity and thermal conductivity smooth the fluid is so small compared to the scales of the flow in many (astrophysical) applications that a mathematical representation through discontinuities is accurate.

The possible presence of discontinuities is also the reason why we use a form of the equations as the basis for numerical integration which differs from HSW1: there exist infinitely many forms of the equations of hydrodynamics that are *mathematically* equivalent to Eq. (2.8). The form chosen here is special in the sense that it has *no derivatives of dependent fluid variables in the source term*. Thus the source term of this form will remain finite everywhere even if discontinuities are present in the solution. This is of course an advantage for numerical integration. This form is uniquely determined in the sense that the combinations in which the *unknowns* enter are fixed. The state of this special form is called the 'conserved state'. As we shall see below, grouping the equations with respect to partial derivatives is essential for the central element in the present method: the Roe solver.

## 3. Numerical approximation and operator splitting

In this section some of the numerical concepts that are important for what follows are introduced. There exists a vast amount of literature on hydrodynamics (e.g. Courant & Friedrichs 1967) and on how to find a numerical approximation to solutions (e.g. HSW2; Roe 1986a). For completeness, and to show how the equations of general relativistic hydrodynamics fit into the theory, some of the numerical aspects of the problem are mentioned briefly.

The most important difficulty of solving hyperbolic partial differential equations like Eq. (2.8) is that they allow discontinuous, "weak" solutions. In the presence of discontinuities the partial differential equation is ill defined, because two or more of the partial derivatives are infinite at a discontinuity in such a way that they cancel. Therefore one needs to take a step backwards and examine the physical situation that led to Eq. (2.8).

Consider a volume $\Delta x^1 \Delta x^2 \Delta x^3$ centred around $(x^1, x^2, x^3)$ in space. The total mass, momentum and energy in this volume change in a time $\Delta x^0$ due to transport across the volume's boundaries and due to the source term:

$$\int \left[ F^0(x^0 + \Delta x^0, x^1, x^2, x^3) \right.$$
$$\left. - F^0(x^0, x^1, x^2, x^3) \right] dx^1 \, dx^2 \, dx^3 +$$
$$\int \left[ F^1(x^0, x^1 + \tfrac{1}{2}\Delta x^1, x^2, x^3) \right.$$
$$\left. - F^1(x^0, x^1 - \tfrac{1}{2}\Delta x^1, x^2, x^3) \right] dx^0 \, dx^2 \, dx^3 +$$
$$\int \left[ F^2(x^0, x^1, x^2 + \tfrac{1}{2}\Delta x^2, x^3) \right.$$
$$\left. - F^2(x^0, x^1, x^2 - \tfrac{1}{2}\Delta x^2, x^3) \right] dx^0 \, dx^1 \, dx^3 +$$
$$\int \left[ F^3(x^0, x^1, x^2, x^3 + \tfrac{1}{2}\Delta x^3) \right.$$
$$\left. - F^3(x^0, x^1, x^2, x^3 - \tfrac{1}{2}\Delta x^3) \right] dx^0 \, dx^1 \, dx^2 -$$
$$\int \mathcal{S} \, dx^0 \, dx^1 \, dx^2 \, dx^3 = 0. \tag{3.1}$$

This more fundamental equation may also be found by integrating Eq. (2.8) over spacetime. In this form the equation remains valid even in the presence of discontinuities. This suggests the following numerical approximation: cover a region of spacetime by a grid of volumes whose centres are fixed in time. For simplicity, the grid is taken to be equidistant in the coordinates $([x^\alpha]_k = [x^\alpha]_0 + k\Delta x^\alpha)$ in this paper. Then for the cell averages $\langle F^0 \rangle$ of the state one has the following second order accurate approximation:

$$[\langle F^0 \rangle]_{i,j,k}^{n+1} = [\langle F^0 \rangle]_{i,j,k}^n - \frac{\Delta x^0}{\Delta x^1} \left( [F^1]_{i+\frac{1}{2},j,k}^{n+\frac{1}{2}} - [F^1]_{i-\frac{1}{2},j,k}^{n+\frac{1}{2}} \right)$$
$$- \frac{\Delta x^0}{\Delta x^2} \left( [F^2]_{i,j+\frac{1}{2},k}^{n+\frac{1}{2}} - [F^2]_{i,j-\frac{1}{2},k}^{n+\frac{1}{2}} \right)$$
$$- \frac{\Delta x^0}{\Delta x^3} \left( [F^3]_{i,j,k+\frac{1}{2}}^{n+\frac{1}{2}} - [F^3]_{i,j,k-\frac{1}{2}}^{n+\frac{1}{2}} \right)$$
$$+ \Delta x^0 [\mathcal{S}]_{i,j,k}^{n+\frac{1}{2}}, \tag{3.2}$$

where $Q_{i,j,k}^n$ denotes the *numerical approximation* to the quantity $Q([x^0]_n, [x^1]_i, [x^2]_j, [x^3]_k)$. This formulation has an important advantage as a numerical approximation because, just like Eq. (2.8), it assures that the flux that leaves one cell will enter another. Numerical schemes with this property are called 'conservative'. Equation (3.2) shows also why it is advantageous to have a finite source term. The difference between the solutions of this numerical formulation and the true solution is due to the approximation of the integrals of the flux and the source term.

What remains to be done to obtain a second order accurate numerical (integration) method is to find expressions for the flux and the source term which, in regions of smooth flow, differ from the true flux and source term at the indicated points in spacetime by no more than terms of second order in $\Delta$. Unfortunately, to determine these terms all at once is so complicated that almost all existing more-dimensional numerical hydrodynamics is based on evaluating these terms using one-dimensional simplified equations. This is perhaps the most limiting assumption for more-dimensional flow at present (Roe 1986b). However with the first genuinely more-dimensional methods for CNRH just emerging (Powell & Van Leer 1989), these improvements are not yet available for incorporation in a GRH scheme and are not included in the present paper. Instead the multi-dimensional partial differential equation is broken into one-dimensional parts. To reduce a complicated partial differential equation like Eq. (2.8) to a series of simpler equations one may proceed as follows: rewrite the differential equation in a form where the partial derivative of the state with respect to time equals the sum of two operators $O_1$ and $O_2$:

$$F^0_{,0} = (O_1 + O_2)(F^0). \tag{3.3}$$

A second order accurate approximation to the solution of the original equation at a time step later is then found by a series of second order accurate integrations of the simpler equations. First integrate

$$F^0_{,0} = O_1(F^0) \tag{3.4}$$

over a half time step. Then use the resultant state as a start value for an integration over a full time step of

$$F^0_{,0} = O_2(F^0). \tag{3.5}$$

The result of this integration is then used as a start value for another integration over a half time step of Eq. (3.4). If these equations are still too complicated for second order accurate integration, the above process can be repeated on the newly formed equations. Each next operator works on the result of the previous operator. In this way any complicated equation can be broken into ever simpler parts (Strang 1968).

Here we apply this operator splitting to reduce Eq. (2.8) to a series of equations in one spatial dimension. As a first operator, part of the source term $T$ is split off. Then three spatial operators are constructed that are each a combination of the flux in one direction and of a part of the source term. The sequence in which the operators are applied should be varied per time step to avoid unphysical drifts as much as possible (Strang 1968). Equation (2.8) can thus be reduced to one set of ordinary differential equations and one set of partial differential equations in one spatial dimension, for each spatial direction:

$$F^0_{,0} = T, \tag{3.6}$$
$$F^0_{,0} + F^1_{,1} = X, \tag{3.7}$$
$$F^0_{,0} + F^2_{,2} = Y, \tag{3.8}$$
$$F^0_{,0} + F^3_{,3} = Z. \tag{3.9}$$

Unfortunately the available information is not sufficient to yield a unique splitting of the source term. In principle the source terms may be chosen freely, subject only to the constraint that $T + X + Y + Z = \mathcal{S}$. This is the reason why a four-dimensional equation is much more difficult to solve than four one-dimensional equations. Each splitting of the source term defines a possible second order accurate update. In physical terms, operator splitting as applied above assumes that all information exchange between cells takes place perpendicular to cell interfaces only.

It is advantageous to make as much use of a priori information about the flow as possible to constrain the source term splitting. Which splitting is best depends on what is known about the flow one is trying to simulate. A form that is often used (e.g. HSW2) is

$$T = \mathcal{S},$$
$$X = Y = Z = 0. \qquad (3.10)$$

A special case of such a splitting is an obvious *choice* for applications without source terms ($\mathcal{S} = 0$). In that case one usually chooses $T = X = Y = Z = 0$. Some possibilities are discussed in Sect. 6 and Appendix C.

Because the solution method for the three spatial operators is similar, the solutions of Eq. (3.8) and (3.9) are not shown explicitly in this paper. Their solutions follow from symmetry and the solution method for Eq. (3.7).

If no special use is made of $X$, for example if equation (3.7) is split once more to obtain an equation without source term and another equation without flux term, it is more cost-effective to choose $X = 0$ and $T = \mathcal{S}$. However, below the source term and flux in a particular direction are used in combination. Equation (3.7) may be rewritten in the form

$$F^0_{,0} + \left(F^1 - \int_{x_c^1}^{x^1} X dx^1\right)_{,1} = 0, \qquad (3.11)$$

where $x_c^1$ is an arbitrary fixed $x^1$ value. We consider those special solutions of this equation which obey

$$F^1_{,1} = X, \qquad (3.12)$$

see Sect. 6 and Appendix C.

Analogous to Eq. (3.2), Eq. (3.11) suggests

$$[\langle F^0 \rangle]^{n+1}_i = [\langle F^0 \rangle]^n_i - \frac{\Delta x^0}{\Delta x^1} \left( \left\{[F^1]^{n+\frac{1}{2}}_{i+\frac{1}{2}} - \tfrac{1}{2}\Delta x^1 [X]^{n+\frac{1}{2}}_{i+\frac{1}{2}}\right\} \right.$$
$$\left. - \left\{[F^1]^{n+\frac{1}{2}}_{i-\frac{1}{2}} + \tfrac{1}{2}\Delta x^1 [X]^{n+\frac{1}{2}}_{i-\frac{1}{2}}\right\}\right). \qquad (3.13)$$

In this expression as in the rest of this paper, the non-varying subscripts $(j, k)$ are omitted for ease of notation. The second order accurate approximations to the flux and the source term of Eq. (3.13) and to the source term of Eq. (3.6) are all that remain to be determined.

It follows from Eq. (3.13) that with the possible exception of the very first time step, only the average state is given within each cell. An approximation to the flow within a cell is quite commonly found by assuming that the state is equal to the known average value everywhere in the cell or that it varies linearly.

The assumption about the flow within a cell, given the average state but not any information about the flow in neighbouring cells, is called the *subgrid model* in this paper. It is discussed in Sect. 6. Subgrid models do not take into account the discontinuities that can be present in the global flow, but local discontinuities, due to state differences between adjacent cells, are supposed to lie on cell interfaces whenever the subgrid model is applied. Consequently, an approximation to the interaction of flows in adjacent cells is sought that remains valid even if the flows in the two cells involved differ substantially.

Mathematically, the key property of hyperbolic equations is that local changes to variables make themselves known to their surroundings at a *finite* speed: because of this property, the changes in state in cells not immediately adjacent to a certain interface cannot influence the flux at that interface for some time. Thus if a small enough time step is chosen (the Courant-Friedrichs-Lewy [CFL] condition) the interface flux is independent of the initial condition outside the two cells that share the interface. If initially the state within each of these two cells is stationary, the interface flux is constant in time (if the metric is time independent). This flux can be computed from the condition that initially stationary states exist in the two half spaces on either side of the interface (see Courant & Friedrichs 1967). These states immediately left and right of the cell boundary under consideration are denoted by $F^0_L$ and $F^0_R$ respectively. The above simplified problem is called the Riemann problem. Its usefulness resides in the fact that the initial condition is a stationary solution away from the interface and that if a small enough time step is taken the information of states of cells farther away has not yet reached the cell boundary under consideration.

The possible presence of discontinuities means that approximations to the solution of the Riemann problem based on for example Taylor series do not always yield accurate results. Instead it is crucial that information is drawn from the proper places, because the state may differ drastically from place to place. To make this statement a bit more precise the concept of characteristics is extremely useful: for one-dimensional problems characteristics are the paths in spacetime along which discontinuities of derivatives propagate, and which serve to bound the domains of dependence and influence of a given event, see Fig. 1. For multi-dimensional problems, characteristic surfaces exist whose properties are rather more complicated. Avoidance of these complications is a strong motive for the simplifying assumption that communication takes place normal to cell boundaries. This allows one to work with quasi one-dimensional equations.

The two-dimensional partial differential Eq. (3.11) can be rewritten as a series of ordinary differential equations (the characteristic equations) along the corresponding characteristic (see e.g. Courant & Friedrichs 1967). For ordinary differential equations no discontinuous solutions exist. Therefore the solution of these characteristic equations varies smoothly along the characteristic and interpolation is useful. This property makes char-acteristics extremely apt to describe flow with strong shocks. Schemes that exploit this notion are called characteristic-based schemes.

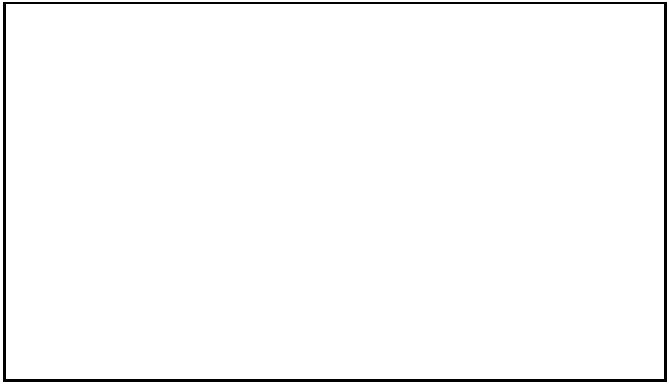

**Fig. 1.** Characteristics in relativistic flow. At the interfaces and at the intermediate time-level in spacetime the three characteristic fluid velocities (forward and backward sound speed relative to the fluid velocity and the fluid velocity itself) and the light cone are drawn. Along each characteristic of the linearized problem a particular combination of dependent variables remains constant. Hence the appropriate interface states may be reconstructed by tracing the characteristics back to the present time. The light cone is the boundary of influence by any macroscopical physical process

With the help of characteristics the Riemann problem can be reduced to a set of nonlinear algebraical equations. However, because the solution is needed only to find the flux in a situation which already was assumed to obey an idealized initial condition (the subgrid model), an easier "approximate Riemann problem" may be solved instead. Two popular methods that solve an approximate Riemann problem are the Roe solver and the ENO scheme (Harten et al. 1987) based on the work of Osher and Salomon (1982). For strong shocks the Roe solver seems marginally better (Chang & Liou 1988). For that reason the present method is based on an extension of the Roe solver.

A Roe solver makes use of approximated characteristics. How this works in detail is discussed in the next section where Roe's original solver (Roe 1981b, 1986a) for CNRH is presented and explained.

## 4. Cartesian non relativistic Roe solver

Roe's (1981b) original CNR solver is the basis for the extensions in this paper. It is applicable to the classical Euler equation in Cartesian coordinates. These equations are obtained by taking the CNR limit of Eq. (2.8), subtracting the energy corresponding to the rest mass in the usual way. They have the form (2.8) with $\mathcal{S} = 0$ and

$$
\begin{aligned}
F^0 &= \left(\rho, \rho u^1, \rho u^2, \rho u^3, \rho e_c\right), \\
F^1 &= \left(\rho u^1, \rho u^1 u^1 + p, \rho u^1 u^2, \rho u^1 u^3, \rho h_c u^1\right), \\
F^2 &= \left(\rho u^2, \rho u^2 u^1, \rho u^2 u^2 + p, \rho u^2 u^3, \rho h_c u^2\right), \\
F^3 &= \left(\rho u^3, \rho u^3 u^1, \rho u^3 u^2, \rho u^3 u^3 + p, \rho h_c u^3\right).
\end{aligned} \quad (4.1)
$$

Here $u^1$, $u^2$, $u^3$ are the velocities in the three spatial directions, while $e_c$ and $h_c$ are the classical (i.e. excluding rest mass energy) energy density of the fluid and the classical enthalpy

$$
\begin{aligned}
e_c &\equiv \tfrac{1}{2}\left[(u^1)^2 + (u^2)^2 + (u^3)^2\right] + \frac{1}{\Gamma - 1}\frac{p}{\rho}, \\
h_c &\equiv \tfrac{1}{2}\left[(u^1)^2 + (u^2)^2 + (u^3)^2\right] + \frac{\Gamma}{\Gamma - 1}\frac{p}{\rho}.
\end{aligned} \quad (4.2)
$$

In the derivation of the equations of Sect. 3 no use was made of particular expressions for the state or the flux. Hence a numerical approximation to the update for the state may be found from Eq. (3.13), provided an expression for the interface flux at the intermediate time is found. This interface flux *does* of course depend on the particular expressions for the state and the flux.

The central ideas of Roe's method derive from the Riemann problem for a set of linear advection equations:

$$W_{,0} + G_{,1} = W_{,0} + \mathbf{A} W_{,1} = 0, \quad (4.3)$$

where $\mathbf{A}$ is a constant matrix. Each of the objects $W$ and $G$ is a five-dimensional vector analogous to the $F^\alpha$ used above.

Let $\lambda_k$ be the eigenvalues of $\mathbf{A}$ and $e_k$ the corresponding right eigenvectors. The set of equations (4.3) can be decoupled by projection onto these eigenvectors, i.e., by determining the coefficients $a_k$ such that

$$W = \sum a_k e_k, \quad (4.4)$$

or equivalently the coefficients $b_k$ such that

$$G = \mathbf{A} W = \sum b_k e_k, \quad (4.5)$$

which are related by

$$b_k = \lambda_k a_k. \quad (4.6)$$

The characteristic form of the set of equations is then

$$db_j = 0 \text{ along } dx^1 = \lambda_j dx^0. \quad (4.7)$$

The eigenvectors represent the different physical waves in a fluid, while the coefficients are the wave strengths (Riemann invariants) and the eigenvalues are the characteristic velocities of the waves. The eigenvalue $\lambda_j$ may thus be identified with the $j^{th}$ characteristic velocity. Equation (4.7) expresses the fact that the corresponding component of the flux does not change along a characteristic in this linear case. The flux at the boundary half a time step later can thus be found by tracing the characteristics back in time from that future point in spacetime (where the flux is needed) to the present time. If the time step obeys the CFL condition, this characteristic intersects the present time within one of the two neighbouring cells:

$$G^{n+\frac{1}{2}}_{i+\frac{1}{2}} = \sum b_k([x^1]_{i+\frac{1}{2}} - \tfrac{1}{2}\lambda_k \Delta x^0, [x^0]_n) e_k. \quad (4.8)$$

Two possible ways to estimate the coefficients at the current time are step- and linear interpolation (see Sect. 9 on which choice should be preferred in a particular situation):

$$b_k([x^1]_{i+\frac{1}{2}} - \tfrac{1}{2}\lambda_k \Delta x^0, [x^0]_n) \approx$$
$$\tfrac{1}{2}[(1+\sigma_k)b_k]_i^n + \tfrac{1}{2}[(1-\sigma_k)b_k]_{i+1}^n, \quad (4.9)$$

or

$$b_k([x^1]_{i+\frac{1}{2}} - \tfrac{1}{2}\lambda_k \Delta x^0, [x^0]_n) \approx$$
$$\tfrac{1}{2}[(1+\nu_k)b_k]_i^n + \tfrac{1}{2}[(1-\nu_k)b_k]_{i+1}^n, \quad (4.10)$$

where $\nu_k = \lambda_k \Delta x^0/\Delta x^1$ and $\sigma_k = \text{sign}(\nu_k)$. In general one may write

$$b_k([x^1]_{i+\frac{1}{2}} - \tfrac{1}{2}\lambda_k \Delta x^0, [x^0]_n) \approx$$
$$\tfrac{1}{2}[\{1 + \sigma_k - (\sigma_k - \nu_k)\psi_k\} b_k]_i^n$$
$$+ \tfrac{1}{2}[\{1 - \sigma_k + (\sigma_k - \nu_k)\psi_k\} b_k]_{i+1}^n, \quad (4.11)$$

where $\psi$ is called the flux limiter (see Sect. 9). Then if

$$W_{i+1}^n - W_i^n = \sum [a_k e_k]_{i+\frac{1}{2}}^n, \quad (4.12)$$

or equivalently

$$G_{i+1}^n - G_i^n = \sum [b_k e_k]_{i+\frac{1}{2}}^n, \quad (4.13)$$

this interpolation yields

$$G_{i+\frac{1}{2}}^{n+\frac{1}{2}} = \tfrac{1}{2}(G_i^n + G_{i+1}^n)$$
$$- \tfrac{1}{2}\sum \left[ \{\sigma_k - (\sigma_k - \nu_k)\psi_k\} b_k e_k \right]_{i+\frac{1}{2}}^n. \quad (4.14)$$

The update of the state is, according to Eq. (3.13),

$$W_i^{n+1} = W_i^n - \tfrac{1}{2}\frac{\Delta x^0}{\Delta x^1} \Big($$
$$\sum \{[1 - \sigma_k + (\sigma_k - \nu_k)\psi_k] b_k e_k\}_{i+\frac{1}{2}}^n$$
$$+ \sum \{[1 + \sigma_k - (\sigma_k - \nu_k)\psi_k] b_k e_k\}_{i-\frac{1}{2}}^n \Big). \quad (4.15)$$

Notice that just like the exact update this numerical update depends only on flux differences, and not on the fluxes themselves!

In order for this method to be of practical use one must be able to deal with nonlinear conservation equations. For this purpose the nonlinear equations are written in the form of Eq. (4.3), where $\mathbf{A}(W)$ is no longer a constant matrix. The method can then be generalized if one can approximate $\mathbf{A}$ by some constant matrix $\tilde{\mathbf{A}}$. Roe (1981b) has formulated the properties that are important for $\tilde{\mathbf{A}}$:

1. It constitutes a linear mapping from the vector space $W$ to the vector space $G$.
2. As $W_L \to W_R \to W$ : $\tilde{\mathbf{A}}(W_L, W_R) \to \mathbf{A}(W)$, where $\mathbf{A} \equiv \partial G/\partial W$.
3. For any $W_L, W_R$ : $\tilde{\mathbf{A}}(W_R, W_L)(W_R - W_L) = G_R - G_L$.
4. The eigenvectors of $\tilde{\mathbf{A}}$ are linearly independent.

Condition (1) is necessary to simplify the original equations to a linear set. Condition (2) is necessary to recover smoothly the linearized algorithm from the nonlinear one. Conditions (3) and (4) were proven to be necessary and sufficient for a conservative scheme with correct jump conditions across a shock (Roe 1981a).

In practice property (4) can be checked afterwards. Matrices $\tilde{\mathbf{A}}$ with properties (1), (2), and (3) can be constructed in several ways (Roe 1981b). The method that was used by Roe for CNRH and that is used in this paper for GRH is based on the exact form

$$\Delta(pq) = \langle p \rangle \Delta q + \langle q \rangle \Delta p, \quad (4.16)$$

where $p$ and $q$ are any quantities, $\Delta p$ is the difference in quantity $p$, and $\langle \rangle$ denotes the arithmetic mean.

Once such a matrix $\tilde{\mathbf{A}}$ is constructed for the particular set of equations, its eigenvectors and eigenvalues may be determined. The projection coefficients are obtained by projecting the state difference (4.12) or equivalently the flux (4.13) on the eigenvectors. The flux and the state update are then given by Eqs. (4.14) and (4.15) respectively.

The Rankine-Hugoniot jump condition for *exact* solutions states that

$$G_R - G_L = \lambda(W_R - W_L), \quad (4.17)$$

where $\lambda$ is the propagation velocity of a stable discontinuity. The key point of Roe solvers is that properties (3) and (4) guarantee that the Riemann problem is linearized in such a way that this correct propagation velocity is found if the initial discontinuity is a pure contact discontinuity or a pure shock, even for nonlinear problems. By comparing Eq. (4.17) to Eqs. (4.13) and (4.12) it may be deduced that $\lambda_j = \lambda$ if only one $a_j \neq 0$. Thus for a discontinuity of only one Riemann invariant Roe's method finds the correct characteristic speed. This property is also the reason why the Roe solver needs the conservative form of the equations, i.e. the form of Eq. (2.8): the exact jump conditions only hold for this form.

By expressing the state and the flux in terms of a parameter vector $K_C(1, h_c, u^1, u^2, u^3)$, where $K_C \equiv \sqrt{\rho}$, and by applying Eq. (4.16), Roe (1981b) found the following explicit expressions for CNRH. The eigenvectors are

$$e_1 = \left(1, \tilde{h}_c - s_c v_c^1, v_c^1 - s_c, v_c^2, v_c^3\right),$$
$$e_2 = \left(1, \tilde{h}_c + s_c v_c^1, v_c^1 + s_c, v_c^2, v_c^3\right),$$
$$e_3 = \left(1, \tfrac{1}{2}\left[(v_c^1)^2 + (v_c^2)^2 + (v_c^3)^2\right], v_c^1, v_c^2, v_c^3\right),$$
$$e_4 = \left(0, v_c^2, 0, 1, 0\right),$$
$$e_5 = \left(0, v_c^3, 0, 0, 1\right), \quad (4.18)$$

where

$$s_c^2 = (\Gamma - 1)\left\{\tilde{h}_c - \tfrac{1}{2}\left[(v_c^1)^2 + (v_c^2)^2 + (v_c^3)^2\right]\right\},$$

$$v_c^i = \frac{\langle K_C u^i \rangle}{\langle K_C \rangle},$$

$$\tilde{h}_c = \frac{\langle K_C h_c \rangle}{\langle K_C \rangle}. \tag{4.19}$$

The characteristic velocities are

$$\lambda_1 = v_c^1 - s_c,$$
$$\lambda_2 = v_c^1 + s_c,$$
$$\lambda_3 = \lambda_4 = \lambda_5 = v_c^1. \tag{4.20}$$

The projection coefficients of a state difference $F_R^0 - F_L^0 \equiv (\delta, \Delta^e, \Delta^1, \Delta^2, \Delta^3)$ are

$$a_1 = \frac{\Gamma - 1}{2s_c^2}\left\{\tfrac{1}{2}\left[(v_c^1)^2 + (v_c^2)^2 + (v_c^3)^2\right]\delta \right.$$
$$\left. - \left[v_c^1\Delta^1 + v_c^2\Delta^2 + v_c^3\Delta^3 - \Delta^e\right]\right\} - \frac{\Delta^1 - v_c^1\delta}{2s_c},$$

$$a_2 = \frac{\Gamma - 1}{2s_c^2}\left\{\tfrac{1}{2}\left[(v_c^1)^2 + (v_c^2)^2 + (v_c^3)^2\right]\delta \right.$$
$$\left. - \left[v_c^1\Delta^1 + v_c^2\Delta^2 + v_c^3\Delta^3 - \Delta^e\right]\right\} + \frac{\Delta^1 - v_c^1\delta}{2s_c},$$

$$a_3 = \frac{\Gamma - 1}{s_c^2}\left(\left\{\tilde{h}_c - \left[(v_c^1)^2 + (v_c^2)^2 + (v_c^3)^2\right]\right\}\delta \right.$$
$$\left. + \left[v_c^1\Delta^1 + v_c^2\Delta^2 + v_c^3\Delta^3 - \Delta^e\right]\right),$$

$$a_4 = \Delta^2 - v_c^2\delta,$$

$$a_5 = \Delta^3 - v_c^3\delta. \tag{4.21}$$

With Eqs. (4.21), (4.20) and (4.18) one can derive numerical approximations to the interface flux (and to the update of the state) from Eqs. (4.14) and (4.15) if $F^0$ is substituted for the state $W$ and $F^1$ for the flux $G$. The flux difference at an interface is found and decomposed under the assumption of a constant state within a cell. Next a correction based on the interface flux difference and the upwind interface flux difference for the relevant characteristic is applied to the flux, by means of the flux limiter $\psi$. This correction models the non-constant flux variation. For nonlinear equations this is not the same as first choosing a suitable flux model and then finding interface states as initial conditions for a Riemann problem, as suggested by e.g. Van Leer (1973, 1974, 1977a, 1977b, 1979).

Unfortunately, a Roe solver does not always yield a physically allowed flux. An example is encountered in the Riemann problem at an interface of two cells filled with a gas with adiabatic exponent $\Gamma = 1.5$ and having $\rho_L = 1, u_L = -2, p_L = 4/3$ and $\rho_R = 4, u_R = 1, p_R = 13/3$ respectively. After calculation the first order upwind flux with the Roe solver it is found that there is no mass flux and yet there is an energy flux. Clearly no physical set of variables exists that can reproduce this Roe flux. An explanation is the fact that the Roe solver represents expansion waves by expansion shocks with one velocity (see Einfeldt et al. 1991). One might consider the use of an exact Riemann solver in cases where the Roe solver cannot cope but including special cases would make the numerical code rather slow. Yet for most problems the Roe solver is remarkably trouble free. Thus a powerful scheme for the classical Euler equations in Cartesian coordinates results.

The above ideas depend on characteristics rather than on the precise functional form of the state and the flux. Hence Roe solvers may be extended to hyperbolic equations other than the CNR Euler equation. Before Eqs. (4.14) and (4.15) can be used to obtain a scheme for the GR Euler equations, the eigenvectors, eigenvalues and projection coefficients for the GR equations have to be found. This is done in Sect. 10. However, first two difficulties need to be addressed. The first of these also appears in SRH and has to do with the relativistic form of the state. It is discussed in Sect. 5. The other difficulty also appears in NRH in non-Cartesian coordinates and is caused by the use of curvilinear coordinates. It is discussed in Sect. 6.

## 5. Determination of the primitive variables

To find an update for the state it is necessary to determine the state or flux differences, the source term and the characteristics of the flow. All of these can be expressed explicitly in terms of the primitive variables $\rho$, $p$, $u^1/u^0$, $u^2/u^0$, $u^3/u^0$. However, only the (average of the) conserved state is known at each time step, as is evident from Eq. (3.13). In RH, in contrast to NRH, the primitive variables are not easily determined from the conserved quantities, because the enthalpy enters the spatial components of the conserved vector in RH. HSW do not encounter this problem because they use another (non conservative) form of the state. The reasons for choosing the conservative form were explained in Sections 2 and 4.

Appendix A is dedicated to a discussion of how to determine the primitive variables given the conserved quantity vector. Three methods are described there: an analytical method, a one-dimensional Newton-Raphson method and a six-dimensional Newton-Kantorovich method. The one-dimensional Newton method was applied to obtain the results in this paper.

## 6. Subgrid model

To predict the flow on a grid of several computational cells one must be able to predict the interaction of the flows in neighbouring cells. This interaction depends on the flow within a cell. In the CNRH scheme discussed in Sect. 4 for example, two interface fluxes each based on information of just one of the two bounding cells are combined by a Roe solver via Eq. (4.14) to construct the desired flux for integration in Eq. (3.13). There are three equations left to solve: (3.6), (3.11) and (3.12). Of these Eq. (3.6) as a whole and Eq. (3.12) (the integration of the source terms) pertain to one cell. Only after these integrations have produced the current interface state on both sides of the interface can the interaction between bounding cells be computed. Hence, the flow is computationally approximated in two distinct steps; the approximation of the flow within a cell and the approximation of the interaction between these flows in neighbouring cells. This section is dedicated to the former step.

The subgrid model describes the assumed behaviour of the primitive variables within a cell as a function of time for Eq. (3.6) and as a function of place for Eq. (3.12).

## 6.1. General considerations

The numerical methods presented in this paper aim to minimize complications due to the metric and source terms in calculating the interactions between adjacent cells by solving physically allowed (approximate) Riemann problems at specific points in spacetime. No complications due to non-physical states or *differences* in the metric enter the Roe solver. To have physically allowed states is important because an approximate Riemann solver may not be expected to yield a physical interface flux at the intermediate time, based on a non-physical current interface state. Any correction that has to be applied to make the final update physical again is arbitrary by its nature. HSW force their results to be physical by adapting $u^0$ ("velocity renormalization", HSW2: Eq. (11)).

The differences in the metric and the source terms are taken into account by subgrid models. These models are used to determine the interface states from the known average state. In principle they can also be used to extrapolate the calculated interface fluxes to the (centres of the) respective cells. Hence the complications due to the metric and source terms are concentrated in the subgrid model and the Roe solver just needs to deal with localized GRH equations. It is anticipated that as a result of this approach the remaining ingredients of the implementation, such as the flux limiter, can stay as they are for CNRH.

In CH without source terms the subgrid model is so simple that it hardly received any attention in Sect. 4. All variables were assumed to be constant within a cell (prior to decomposition onto the eigenvectors). This model is not adequate in more general geometries and/or coordinates. This is especially apparent in curved spacetime: a vector at one point in spacetime is not necessarily a vector at another point, and a linear combination of two vectors that exist at different points of spacetime generally does not yield a vector in a location in between.

Desirable properties of subgrid models are easily identified by considering CNRH without source terms. There the usual *choice* is a subgrid model with constant variables given by the cell-averaged state. This model has four advantages:

1. The subgrid model is computationally cheap. The only work involved in this subgrid model is the determination of the primitive variables that belong to the given averaged state.
2. The average of the state over the cell is equal to the prescribed cell-average.
3. The prescribed state is a stationary solution. This ensures firstly that the interface problem is a Riemann problem and secondly that a one-dimensional stationary solution is preserved.
4. The prescribed interface states are always physically allowed.

In CSRH without source terms the constant variable model retains these advantages and consequently it is the obvious choice. However in case a non-zero source term or non-constant known functions (metric) enter the problem, the four properties are in conflict. For example, a constant state cheaply assures the correct cell-averaged state but is in general not a stationary solution. Hence, some of the desirable properties above have to be sacrificed.

On the one hand one can decide to discard property (3). In that case the constant state model can be accepted. Even then extra calculations are required if the metric varies because the same state will produce different primitive variables and consequently different fluxes at different places within a cell. In addition, this subgrid model cannot guarantee a physical interface state, i.e. property (4) is compromised as well. Hence, the disadvantages of this subgrid model are that it diffuses even a one-dimensional stationary solution and that non-physical interface states cannot be ruled out.

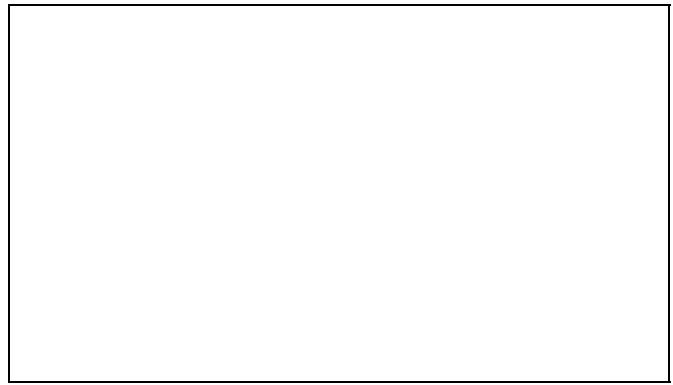

**Fig. 2.** Subgrid model based on extrapolation. The cell average data is assumed to yield the variables at the cell centre. To obtain the variables at the interfaces the variables at the cell centre are extrapolated via an assumption, for example that of one-dimensional stationary flow. These interface variables are subsequently used as input for the (approximate) Riemann problems

On the other hand, to ensure minimal damage to stationary one-dimensional solutions, one may use these as a basis for a subgrid model. These solutions are in general nonlinear functions of the state. Therefore it is no longer feasible to enforce the cell-averaged state to be correct. Instead the cell-averaged state is assumed to be representative of the state at the cell centre. The subgrid model is then fixed by a value rather than by an integral equation as condition to select the proper solution from those permitted by the differential equation. The subgrid model is now effectively an extrapolation of the current data at the cell centre in space or in time. Schematically this extrapolation is indicated in Fig. 2.

Notice that, even though the cell-averaged state may not be correct, such a model yields a conservative method as long as the flux added to a cell is subtracted from a neighbour. A possible, relatively cheap implementation of this kind of subgrid model is to use the primitive variables that belong to the average state to define a constant average source term for the flux, i.e.

$$[F^1]_{i+\frac{1}{2}-}^{n+\frac{1}{2}-} = [F^1]_i^{n+\frac{1}{2}-} + \tfrac{1}{2}\Delta x^1 X_i^{n+\frac{1}{2}-}. \tag{6.1}$$

This subgrid model preserves one-dimensional stationary solutions to second order, but only has a first order correct cell-averaged state, and does not ensure a physical interface state.

Hence, both subgrid models above cannot guarantee a physically allowed interface state. In GR, parallel transport is the physical way to transport vectors from one point to another. Indeed, parallel transport is a possible subgrid model that yields physical interface states. However parallel transport makes no use of the symmetries of the flow that usually underlie the choice for non-Cartesian grids. This can easily be seen for curvilinear coordinates in flat spacetime: in that case parallel transport leaves vectors constant along straight lines, independent of the symmetry of the coordinates. As a result a radial velocity vector is no longer radial after parallel transport in any other direction than the radial direction, even if spherical coordinates are used.

The transport presented below yields physically allowed interface states and makes use of symmetries of the flow with respect to the chosen coordinates. The freedom inherent in the source term splitting is used to choose a special case that is based on one-dimensional stationary solutions. As a result the one-dimensional Eqs. (3.6) and (3.12) can be integrated analytically.

### 6.2. Algebraic extrapolation

Both the extrapolation in space and in time may be done algebraically if (quasi) one-dimensional flow in curved spacetime is assumed, and the source terms are split according to their metric derivative:

$$\begin{aligned}
T &= \sqrt{-g}\left(0, -g^{\alpha\beta}g_{\beta\gamma,0}T^{\gamma 0} + \tfrac{1}{2}g^{\alpha 0}g_{\beta\gamma,0}T^{\beta\gamma}\right), \\
X &= \sqrt{-g}\left(0, -g^{\alpha\beta}g_{\beta\gamma,1}T^{\gamma 1} + \tfrac{1}{2}g^{\alpha 1}g_{\beta\gamma,1}T^{\beta\gamma}\right), \\
Y &= \sqrt{-g}\left(0, -g^{\alpha\beta}g_{\beta\gamma,2}T^{\gamma 2} + \tfrac{1}{2}g^{\alpha 2}g_{\beta\gamma,2}T^{\beta\gamma}\right), \\
Z &= \sqrt{-g}\left(0, -g^{\alpha\beta}g_{\beta\gamma,3}T^{\gamma 3} + \tfrac{1}{2}g^{\alpha 3}g_{\beta\gamma,3}T^{\beta\gamma}\right).
\end{aligned} \tag{6.2}$$

In the spatial directions this extrapolation will henceforth be called stationary extrapolation, because it is the correct subgrid model for one-dimensional stationary flow (i.e. the state and the metric are time independent). Applied in the time direction it is called 'homogeneous extrapolation', because it is correct for one-dimensional homogeneous flows (i.e. flows in which the flux and the metric are independent of the spatial coordinates). This splitting has the additional advantage that $T = 0$ if $g_{,0} = 0$ and $X = 0$ if $g_{,1} = 0$. Thus if the metric is time independent the time extrapolation can be skipped, and if the metric is homogeneous the space extrapolation can be skipped.

More important than this last property is the fact that if analytical extrapolation is used the metric is taken into account exactly. On a Cartesian grid an unphysical prediction occasionally happens due to approximations in the Riemann solver (see Sect. 4) or due to the flux limiter. In GRH however, not taking the metric exactly into account is an additional very important source of these errors. Near a black hole for instance, some components of the metric tend to infinity. Consequently an arbitrarily small approximation error in the source term can dominate the flow close to the horizon, often resulting in the erroneous prediction of states that cannot physically exist.

We now derive the equations for integration in the 1-direction. Extrapolation in the other directions is similar. By assuming that while integrating in one direction (here the 1-direction) all functions are constant in all other directions, it follows from Eq. (3.11) that for a stationary solution the mass flux is equal to a constant $D$:

$$\sqrt{-g}\rho u^1 = D. \tag{6.3}$$

Replacing the Christoffel symbols by their definition Eq. (2.7) and taking only the 1-derivatives into account for the extrapolation in the 1-direction yields

$$\begin{aligned}
(\sqrt{-g}T^1{}_\epsilon)_{,1} &= (\sqrt{-g}g_{\epsilon\alpha}T^{1\alpha})_{,1} \\
&= g_{\epsilon\alpha,1}\sqrt{-g}T^{1\alpha} + g_{\epsilon\alpha}(\sqrt{-g}T^{1\alpha})_{,1} \\
&= g_{\epsilon\alpha,1}\sqrt{-g}T^{1\alpha} - g_{\epsilon\alpha}g^{\alpha\delta}g_{\gamma\delta,1}\sqrt{-g}T^{1\gamma} \\
&\quad + \tfrac{1}{2}g_{\epsilon\alpha}g^{\alpha 1}g_{\gamma\delta,1}\sqrt{-g}T^{\gamma\delta} \\
&= g_{\epsilon\alpha,1}\sqrt{-g}T^{1\alpha} - g_{\epsilon\gamma,1}\sqrt{-g}T^{1\gamma} = 0, \quad \epsilon = 0, 2, 3.
\end{aligned} \tag{6.4}$$

Hence $\sqrt{-g}T^1{}_\epsilon$ is constant. In combination with Eq. (6.3) this implies for $D \neq 0$ that

$$(hu_\epsilon)_{,1} = 0, \quad \epsilon = 0, 2, 3, \tag{6.5}$$

If $D = 0$ this relation is adopted. For $\epsilon = 1$ one has

$$\begin{aligned}
(\sqrt{-g}T^1{}_1)_{,1} &= \tfrac{1}{2}g_{\gamma\delta,1}\sqrt{-g}T^{\gamma\delta} = -\tfrac{1}{2}g^{\gamma\delta}{}_{,1}\sqrt{-g}T_{\gamma\delta} \\
&\quad - \tfrac{1}{2}\sqrt{-g}\frac{\rho}{h}\left[(g^{\gamma\delta}hu_\gamma hu_\delta)_{,1} - g^{\gamma\delta}(hu_\gamma hu_\delta)_{,1}\right] \\
&\quad - \tfrac{1}{2}g^{\gamma\delta}{}_{,1}\sqrt{-g}pg_{\gamma\delta} \\
&= \sqrt{-g}\rho h_{,1} + \sqrt{-g}\rho u^1(hu_1)_{,1} + (\sqrt{-g})_{,1}p, \tag{6.6}
\end{aligned}$$

where in the last step the identity

$$\tfrac{1}{2}g_{\gamma\delta}g^{\gamma\delta}{}_{,1} = -\frac{1}{\sqrt{-g}}(\sqrt{-g})_{,1} \tag{6.7}$$

is used. Because the mass flux $D$ is constant, one may also write

$$\begin{aligned}
(\sqrt{-g}T^1{}_1)_{,1} &= (Dhu_1)_{,1} + (\sqrt{-g}p)_{,1} \\
&= \sqrt{-g}\rho u^1(hu_1)_{,1} + (\sqrt{-g}p)_{,1}. \tag{6.8}
\end{aligned}$$

Equating the two expressions (6.6) and (6.8) for $\sqrt{-g}T^1{}_1$, one finds

$$(\ln p - \Gamma \ln \rho)_{,1} = 0, \tag{6.9}$$

and hence

$$p = \kappa \rho^\Gamma, \tag{6.10}$$

where $\kappa$ is a constant. This relation is the requirement that entropy should remain constant in stationary interpolation (shocks are supposed to lie on cell boundaries only). This means that this subgrid model prescribes isentropic flow. This is not the case for an arbitrary splitting of the source terms. The entropy is constant for the splitting used here because it *systematically* ignores all changes in other directions than the direction of extrapolation.

The five relations (6.3), (6.5) and (6.10) yield a set of algebraic equations for the stationary solution of the (quasi) one-dimensional flow. To find the primitive variables at an arbitrary location, one can combine these relations into an implicit equation for $\rho$:

$$q(\rho) \equiv \frac{\rho\sqrt{h_0^2 - h^2}}{h} = \frac{|D|}{\sqrt{-g}\sqrt{g^{11}}} \equiv Q. \qquad (6.11)$$

Here $h_0$ is the value of the enthalpy if $u^1 = 0$ (or equivalently $D = 0$)

$$h_0^2 \equiv \frac{(g^{1\epsilon} h u_\epsilon)^2}{g^{11}} - g^{\epsilon\eta} h u_\epsilon h u_\eta, \qquad (6.12)$$

where the summation only involves $\epsilon, \eta = 0, 2, 3$. The metric is evaluated at the point where the primitive variables are to be found. Physical solutions should have $1 < h \leq h_0$. Given the density, the other primitive variables follow simply from the constants of the flow as

$$p = \kappa \rho^\Gamma, \qquad (6.13)$$

$$u^1 = \frac{D}{\sqrt{-g}\rho}, \qquad (6.14)$$

$$u_\epsilon = \frac{h u_\epsilon}{1 + \frac{\Gamma}{\Gamma - 1}\kappa \rho^{\Gamma - 1}}. \qquad (6.15)$$

### 6.3. Solution of the algebraic equations

The constants $\kappa, D, hu_0, hu_2$, and $hu_3$ are determined by assuming that the cell averages are representative for the cell centre. Next, Eq. (6.11) can be solved for $\rho$, by a Newton method for instance. The solution is the intersection of the left-hand side function $q$, that is zero at $\rho = 0$ and at $\rho_0 > 0$ and that has a negative second derivative in that interval, and the constant on the right-hand side. Hence there are either zero, one, or two roots. It is shown below that when two roots exist one corresponds to a supersonic and the other to a subsonic velocity. In case one root exists the flow is sonic. This situation is similar to the well studied problem of classical steady flow along a streamline: for a given mass flux, energy flux and entropy there are either zero, one, or two possible states. In case there are two possible states one corresponds to a subsonic and one to a supersonic solution.

The necessary and sufficient condition for the existence of roots may be found by examining $q$. The maximal mass flux that can be reached for given $hu_\epsilon$, $\epsilon = 0, 2, 3$ and $\kappa$ is found by considering the $\rho$ derivative of $q$:

$$\frac{dq}{d\rho} = \frac{P(h)}{h^2\sqrt{h_0^2 - h^2}}, \qquad (6.16)$$

where

$$P(h) = -h^3 + (2 - \Gamma)h_0^2 h + h_0^2(\Gamma - 1), \qquad (6.17)$$

and we used the relation

$$\rho \frac{dh}{d\rho} = (\Gamma - 1)(h - 1). \qquad (6.18)$$

The polynomial $P(h)$ has at least one root $h_*$ in the range $1 < h_* < h_0$, because $P(1) = h_0^2 - 1 > 0$ and $P(h_0) = -(\Gamma - 1)h_0^2(h_0 - 1) < 0$. In fact there is only one root in the physically allowed region because the $h$ derivative of $P$ changes sign at $h_\pm = \pm\sqrt{(2 - \Gamma)/3}h_0$. This means that $P$ has one extremum for positive $h$. Thus precisely one root $h_*$ exists that has $1 < h_* < h_0$ (other roots have $h < 0$, if they exist). The root formula for cubic equations shows there are one, two or three roots in total, respectively if $h_0/h_1 < 1$, $h_0/h_1 = 1$ or $h_0/h_1 > 1$, where

$$h_1 = \frac{3(\Gamma - 1)}{2(2 - \Gamma)}\sqrt{\frac{3}{2 - \Gamma}}. \qquad (6.19)$$

It is computationally faster not to use the cubic root formula to find the physical root but rather a Newton method. A good starting value for the Newton method is $h = h_0$. For that value $PP'' > 0$ and both $P'$ and $P''$ do not change sign in the interval $(h_*, h_0)$, so the root is approached from one side (here $h > h_*$), see Abramowitz and Stegun (1970). Hence one always finds the required largest root $h_*$. Given this $h_*$ the corresponding density may be found from

$$\rho_* = \left[\frac{\Gamma - 1}{\Gamma \kappa}(h_* - 1)\right]^{\frac{1}{\Gamma - 1}}. \qquad (6.20)$$

Then the maximal mass flux that can be supported by $hu_\epsilon$, $\epsilon = 0, 2, 3$ and $\kappa$ is

$$|D_*| = \sqrt{-g}\sqrt{g^{11}}\rho_* \frac{\sqrt{h_0^2 - h_*^2}}{h_*}. \qquad (6.21)$$

If the required mass flux $|D| > |D_*|$ then no solution exists; if $|D| = |D_*|$ one solution exists that has precisely the sonic velocity (see below), and if $|D| < |D_*|$ both a supersonic and a subsonic solution exist. These solutions are found by applying a Newton method to Eq. (6.11) with start value $\rho = 0$ and $\rho = \rho_{max}$ respectively, where $\rho_{max}$ is given by

$$h_{max} \equiv 1 + \frac{\Gamma}{\Gamma - 1}\kappa \rho_{max}^{\Gamma - 1} = \frac{h_0}{\sqrt{1 + \left(\frac{Q}{\rho_0}\right)^2}}. \qquad (6.22)$$

By examining the relevant derivatives it may verified that both solutions are approached from one side with these initial values. The simpler starting value given by $h = h_0$ cannot be used because $dq/d\rho$ is infinite for $\rho = \rho_0$, see Eq. (6.16). This would cause the Newton method not to update the approximation to $\rho$, even though Eq. (6.11) is not obeyed. The starting value $\rho_{max}$ is good because it is smaller than $\rho_0$ and greater than or equal to the largest root of Eq. (6.11), it has $q(\rho_{max}) < Q$ and it has a well defined $dq/d\rho$ for $Q \neq 0$, while for $Q = 0$ it is the correct solution.

When no solution exists, one has to conclude that the chosen subgrid model is false. In GRH there can be two physical reasons for the absence of solutions. In flow around a black hole, for example, the fluid may not have sufficient energy to reach a certain larger distance from the hole. This is expressed by the failure of the interpolation procedure because $h_0 < 1$. To force a solution when $h_0 \leq 1$, $h_0$ is multiplied by the ratio $(1+\epsilon)/h_0$ to yield $h_0 = 1+\epsilon$, with $\epsilon$ a small positive number. This will yield very small interface fluxes. To avoid inconsistencies between the interface quantities and the central quantities, the central state is also replaced by a state with the original $D$, $\kappa$ and $u_\epsilon$, $\epsilon = 0, 2, 3$ but with an enthalpy that is multiplied by the same ratio. Small changes to the state can be explained away because it was assumed that the cell-average of the state is the actual state at the cell centre.

The other possible cause is that the cell's surface area is too small for a fluid of a fixed energy to transport the required mass flux. In that case $h_0 > 1$, but the mass flux $D$ is too large to allow a solution. The easiest way to find an extrapolation anyway is to put in a sonic solution. Numerical experimenting showed that this can best be done by adapting the mass flow rate to $|D| := |D_*|$. Again to avoid inconsistencies between the extrapolated and the original flows, the original state is replaced by a state with the original $\kappa$ and $hu_\epsilon$, $\epsilon = 0, 2, 3$ but with $|D| = |D_*|$. The subsonic branch is chosen for the corresponding state if it is upwind of the interface and the supersonic branch if it is downwind. Thus the existence issue is settled.

Two solutions exist in nearly all cases encountered in practice. Equation (6.11) may be rewritten in terms of velocity as

$$\left(\frac{h}{h_0}\right)^2 = \frac{g^{11}}{u^1 u^1 + g^{11}}. \tag{6.23}$$

Since the speed of sound $a$ is given by

$$a^2 \equiv \frac{\Gamma p}{\rho h} = \frac{(\Gamma - 1)(h - 1)}{h}, \tag{6.24}$$

the Mach number of the flow is

$$M^2 \equiv \frac{1}{a^2} \frac{u^1 u^1}{(u^1 u^1 + g^{11})} = \frac{h}{(\Gamma - 1)(h - 1)}\left[1 - \left(\frac{h}{h_0}\right)^2\right]. \tag{6.25}$$

By comparing this expression to the polynomial $P$ above it may be seen that $M^2 > 1$ if $P > 0$, i.e. if $\rho < \rho_*$, and $M^2 < 1$ if $P < 0$, i.e. if $\rho > \rho_*$. Hence if two solutions of Eq. (6.17) exist, one is subsonic and the other supersonic. The selection of the proper root is the remaining issue.

The choice between the sub- and supersonic solution requires attention. It does not suffice to adhere to the same branch as the central state in all cases. Under the influence of a source term, smooth flow can go from a subsonic to a supersonic regime through a sonic point. A well-known example of such a smooth transition occurs in the de Laval nozzle. Therefore, one must check whether such a sonic point is present in the flow within a cell. This is done by solving $D_*(x^1) = D$. After a starting value $x^1$ is chosen, $D_*$ is determined by solving Eq. (6.17), (6.20) and (6.21) for that $x^1$. Next $x^1$ is updated by a Newton method:

$$x^1 := x^1 - \frac{D_* - D}{D_{*,1}}, \tag{6.26}$$

where

$$D_{*,1} = \frac{D_*}{\sqrt{-g}\sqrt{g^{11}}}\left(\sqrt{-g}\sqrt{g^{11}}\right)_{,1} + \frac{1}{2}\frac{D_*}{h_0^2 - h_*^2}\left(h_0^2\right)_{,1}, \tag{6.27}$$

with

$$\left(h_0^2\right)_{,1} = \frac{2(g^{1\epsilon}hu_\epsilon)(g^{1\eta}_{,1}hu_\eta)}{g^{11}} - \frac{(g^{1\epsilon}hu_\epsilon)^2}{(g^{11})^2}g^{11}_{,1} - g^{\epsilon\eta}_{,1}hu_\epsilon hu_\eta. \tag{6.28}$$

If a solution $x^1$ is found which lies between the cell centre and its interface, the solution is assumed to switch branches. This procedure is not very sensitive to numerical accuracy, because flow in the neighbourhood of a point with flow on one branch is on the same branch unless the flow is almost sonic. In that case the difference between the two branches is small. In fact, accidentally picking the wrong branch every once in a while may be thought of as a stability check on the flow: if it makes a considerable difference, the flow pattern one is trying to approximate probably is not stable.

### 6.4. Disadvantages of stationary extrapolation

Stationary extrapolation also has two disadvantages. The first is of a general nature and the second is only a disadvantage if extrapolation of a flux is required for which the corresponding primitive variables are not known. Such a situation occurs if an interface Roe flux is to be extrapolated to the cell centre.

The first disadvantage is related to a typical superresolution problem. An assumption is used to get more resolution than the grid actually contains. In case that assumption is incorrect a penalty is paid. A specific example is given in Appendix B.

The second disadvantage is that for algebraic extrapolation the primitive variables must be known. The Roe solver only yields the upwind flux and not the upwind state (see Sect. 4) at the interface of two cells. The primitive variables can be determined uniquely from the state, but in general not from the flux. This problem is discussed in more detail in Appendix C.

To avoid having to determine the primitive variables from the flux we use the same source term that was used to get information from the cell centre to the interface to transport the flux in the opposite direction. If

$$[X]_i^{n+\frac{1}{2}-} \equiv \frac{1}{\Delta x^1}\left([F^1]_{i+\frac{1}{2}-}^{n+\frac{1}{2}-} - [F^1]_{i-\frac{1}{2}+}^{n+\frac{1}{2}-}\right), \tag{6.29}$$

then

$$[F^1]_{i+}^{n+\frac{1}{2}+} = [F^1]_{i+\frac{1}{2}}^{n+\frac{1}{2}+} - \frac{1}{2}\Delta x^1 [X]_i^{n+\frac{1}{2}-}, \tag{6.30}$$

and

$$[F^1]_{i-\frac{1}{2}}^{n+\frac{1}{2}+} = [F^1]_{i-\frac{1}{2}}^{n+\frac{1}{2}+} + \frac{1}{2}\Delta x^1 [X]_i^{n+\frac{1}{2}-}. \tag{6.31}$$

As is discussed in appendix C one must add a small correction term to this source term to make the resulting integration of the partial differential equation second order accurate.

*6.5. Concluding remarks*

So far only extrapolation in the spatial direction has been discussed. Most of what has been said is, mutatis mutandis, applicable also to time extrapolation. The problems due to non-uniqueness do not occur, because the primitive variables are uniquely determined by the state. For time independent metrics it is trivial. A schematic overview of all extrapolations involved is shown in Fig. 3.

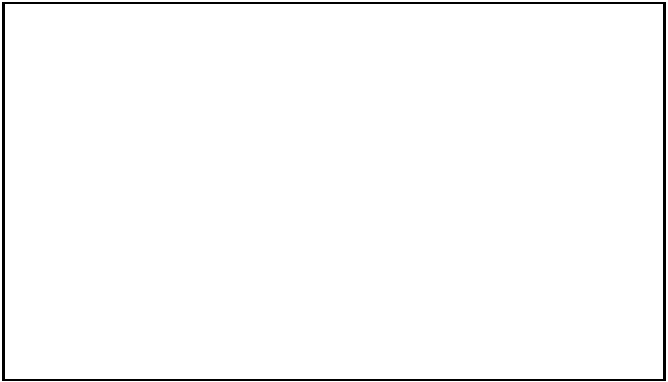

**Fig. 3.** Schematic overview of the extrapolations involved to obtain an update. The arrows indicate extrapolation of information in one point of spacetime to another. The time levels $n + \frac{1}{2}-$ and $n + \frac{1}{2}+$ indicate are just before and just after the approximate Riemann problem at time $n + \frac{1}{2}$, respectively. For time independent metrics the extrapolations in the time direction are trivial

In this section a physical interface state was found by stationary/homogeneous extrapolation of the cell centre data. Hence if an exact GR Riemann solver is applied to the interface states, a physical interface flux is guaranteed. If this interface flux is extrapolated back to the cell centre (the primitive variables are known after an exact Riemann solver) a physical update can be guaranteed in GR.

If a Roe solver is applied, however, a physical update cannot be guaranteed because of two reasons: first, a Roe interface flux is itself sometimes unphysical. This problem is not specific to the extensions discussed in this paper. An example in CNRH was given in Sect. 4. Second, the Roe interface fluxes are not extrapolated back to the cell centre by stationary extrapolation. Instead we used Eqs. (6.31) and (6.30).

Numerical experience shows that with these subgrid models an unphysical update rarely occurs. If it does occur, a new update is calculated by using the flux that corresponds to the cell centre and the flux that is found after extrapolating the upwind state all the way to the cell centre. If this does not improve matters this flux difference is applied for a progressively smaller time step, until a physical update is found. By ensuring that the time step is zero after a finite number of trials a physical update is always found, because if all else fails the new state is kept equal to the present state. Admittedly, this is rather ad hoc.

## 7. Approximate Riemann problem

In this section the interaction of the fluid in adjacent cells is described. As stated is Sect. 3 discontinuities can be present in the flow. The subgrid models presented in Sect. 6 do not take these into account. Hence discontinuities are assumed to lie on cell interfaces. As a result an approximation to the interaction of flows in adjacent cells is sought that remains valid even if the flows in the two cells involved differ substantially.

The equations of GRH are more complicated than the CNRH equations discussed in Sect. 4 in the sense that generally a source term is present and that the state vector and the flux depend on a known continuous metric $g_{\alpha\beta}$. Denote these known continuous functions by $g$. Then one has

$$W(u, g)_{,0} + G(u, g)_{,1} = B(u, g). \tag{7.1}$$

It suffices to find the evolution of the unknowns $u$ since the flux at the cell interface at the intermediate time can be built from combinations of these unknowns and the known quantities. The initial values of the unknowns follow from the initial conserved quantities. It is possible to isolate the differentiated unknowns by rewriting Eq. (7.1) as

$$\frac{\partial W}{\partial u} u_{,0} + \frac{\partial G}{\partial u} u_{,1} = J, \tag{7.2}$$

where

$$J \equiv B - \frac{\partial W}{\partial g} g_{,0} - \frac{\partial G}{\partial g} g_{,1}. \tag{7.3}$$

Let $e_k$ be the right eigenvectors and $\lambda_k$ be the eigenvalues of

$$\mathbf{A} = \frac{\partial G}{\partial u}\left(\frac{\partial W}{\partial u}\right)^{-1}, \tag{7.4}$$

and let

$$\frac{\partial G}{\partial u} du - J dx^1 = \Sigma e_k db_k, \tag{7.5}$$

then the characteristic form of this equation is (cf. Eq.[4.7])

$$db_j = 0 \text{ along } dx^1 = \lambda_j dx^0. \tag{7.6}$$

Notice that if the known functions are time independent, one has

$$\frac{\partial G}{\partial u} du - J dx^1 \equiv dG - B dx^1, \tag{7.7}$$

and Eqs. (7.7) and (7.6) indicate that the interface states determined by stationary extrapolation may be used to determine Roe fluxes at these interfaces. A possible time dependence of the metric is taken into account by Eq. (3.6).

## 8. Implementation for general relativistic flow

In Sections 3, 5, 6, and 7 the ingredients of numerical methods for GRH were discussed. In this section these elements are integrated. The update of the fluid state after a time step $\Delta t$ is determined as follows:

1. Given the state $[F^0]_i^{n-\frac{1}{2}+}$ determine the primitive variables $u_i^{n-\frac{1}{2}+}$.
2. Calculate the spatial source term $X_i^{n-\frac{1}{2}+}$.
3. Extrapolate the primitive variables to the intermediate time level $[x^0]_{n+\frac{1}{2}}$ by homogeneous extrapolation.
4. Calculate the corresponding state $[F^0]_i^{n+\frac{1}{2}-}$ and spatial source term $X_i^{n+\frac{1}{2}-}$.
5. Extrapolate these unknowns from the cell centre $[x^1]_i$ to the interfaces at $[x^1]_{i-\frac{1}{2}}$ and $[x^1]_{i+\frac{1}{2}}$ by stationary extrapolation. The resulting unknowns are denoted by $u_{i-\frac{1}{2}+}^{n+\frac{1}{2}-}$ and $u_{i+\frac{1}{2}-}^{n+\frac{1}{2}-}$ respectively.
6. Calculate the states $[F^0]_{i-\frac{1}{2}+}^{n+\frac{1}{2}-}$ and $[F^0]_{i+\frac{1}{2}-}^{n+\frac{1}{2}-}$ immediately right and left of the cell's interfaces.
7. At each interface (e.g. at $[x^1]_{i-\frac{1}{2}}$) solve the (approximate) Riemann problem defined by the states immediately left and right of that interface using the Roe solver: given the primitive variables $[u]_{i-\frac{1}{2}-}^{n+\frac{1}{2}-}$ and $[u]_{i-\frac{1}{2}+}^{n+\frac{1}{2}-}$ and the metric at the interface $[g]_{i-\frac{1}{2}}^{n+\frac{1}{2}}$, determine the characteristic speeds $\lambda_k$ and the eigenvectors $e_k$ at each interface. Project the state difference on the eigenvectors:

$$[F^0]_{i-\frac{1}{2}+}^{n+\frac{1}{2}-} - [F^0]_{i-\frac{1}{2}-}^{n+\frac{1}{2}-} \equiv \sum [a_k e_k]_{i-\frac{1}{2}}^{n+\frac{1}{2}-}. \tag{8.1}$$

8. Update the intermediate state according to

$$[F^0]_i^{n+\frac{1}{2}+} \equiv [F^0]_i^{n+\frac{1}{2}-} - \frac{1}{2}\frac{\Delta x^0}{\Delta x^1}\Bigg(\\ \sum \left[\{1 - \sigma_k + (\sigma_k - \nu_k)\psi_k\}\lambda_k a_k e_k\right]_{i+\frac{1}{2}}^{n+\frac{1}{2}-} \\ + \sum \left[\{1 + \sigma_k - (\sigma_k - \nu_k)\psi_k\}\lambda_k a_k e_k\right]_{i-\frac{1}{2}}^{n+\frac{1}{2}-}\Bigg) \\ + \Delta x^0 \left(X_i^{n-\frac{1}{2}+} - X_i^{n-\frac{1}{2}-}\right). \tag{8.2}$$

This completes the recipe for the state update. For every newly determined state or flux, new unknowns should be determined. After stationary/homogeneous extrapolation the new density is available and consequently determining the new unknowns is trivial. However after changes in the state other than those due to extrapolation, the unknowns must be determined by a more expensive procedure (see Sect. 5). The total amount of work per cell and per time step in one spatial direction consists of one time extrapolation, two space extrapolations, one approximate Riemann problem, and one primitive variables determination from the state.

The implementation is such that, by virtue of the stationary extrapolation from the cell centre to the interface, a one-dimensional stationary flow in a time independent metric is preserved. As the Roe solver only yields the interface flux, and the primitive variables that need to be known for stationary extrapolation cannot always be determined uniquely from a flux, extrapolation from the interface to the cell centre is avoided. This is the main justification for the seemingly unbalanced procedure of applying stationary extrapolation from the cell centre to the cell interface but not vice versa. See Appendix C for a more elaborate discussion.

Although the source difference term in step (8) is formally required to make the method second order accurate in time, numerical experience shows that in most cases its omission does not cause a significantly different result. Nevertheless it was retained in the calculation of the results presented in this paper.

In case there is no source term and no dependence on known functions the same formalism as in Sect. 4 is automatically recovered, because the extrapolations become trivial. Only steps (1), (7), and (8) of the recipe are then necessary.

In Appendix C several alternative implementations are discussed and the reasons for choosing this particular one are explained.

For smooth flow the method is second order accurate both in space and in time. This is obvious if one realizes that the recipe above consists of a second order accurate integration in time and a second order accurate integration in space combined in such a way as to yield a second order result, see Sect. 3. Alternatively one may prove that the methods are second order accurate by writing out the nested dependencies. This is done in Appendix D.

## 9. Flux Limiter

In this section the algorithm that switches between constant and linear interpolation of the cell averaged data is discussed. It was developed by Roe for the classical case, and we apply it in GRH as well. Because of the different merits of constant and linear interpolation it is desirable to switch between Eqs. (4.9) and (4.10) depending on what seems the best interpretation of the fluid state. Near shocks the jump of conserved variables should be reproduced. Although the true solution may be discontinuous at any point in space, the limited information available per cell generally forces one to place the jump at a cell boundary. Therefore near shocks the average state can be considered to be a characteristic value throughout a cell, whereas in smooth regions of the flow it is better to linearly interpolate between adjacent cells. The general interpolation method can be written as (see Eq.[4.14])

$$[F^1]_{i+\frac{1}{2}}^{n+\frac{1}{2}} = \frac{1}{2}\left([F^1]_{i+\frac{1}{2}+}^{n+\frac{1}{2}-} + [F^1]_{i+\frac{1}{2}-}^{n+\frac{1}{2}-}\right) \\ - \frac{1}{2}\sum \left[\{\sigma_k - (\sigma_k - \nu_k)\psi_k\}\lambda_k a_k e_k\right]_{i+\frac{1}{2}}^{n+\frac{1}{2}-}. \tag{9.1}$$

The actual choice between interpolation assumptions is made by a "flux limiter" function $\psi_k = \psi(r_k)$ of a parameter $r_k$ that should indicate how smooth or discontinuous the flow is. Notice the assumption, implicit in the notation, that for switching purposes the different characteristics can be treated separately.

A convenient way to define the parameter $r_k$ is

$$r_k \equiv \frac{[a_k]^{n+\frac{1}{2}-}_{i+\frac{1}{2}-\sigma_k}}{[a_k]^{n+\frac{1}{2}-}_{i+\frac{1}{2}}}. \tag{9.2}$$

This parameter is approximately equal to unity in regions of smooth flow; consequently, $\psi(1) = 1$ should hold as may be deduced from Eq. (4.11). Numerical experience has shown that "Superbee" is a good choice (Roe 1985):

$$\psi(r) = \begin{cases} 0 & r < 0; \\ \min(1, 2r) & 0 \leq r \leq 1; \\ \min(2, r) & 1 < r. \end{cases} \tag{9.3}$$

In RH the slightly modified form

$$\psi(r) = \begin{cases} 0 & r < 0; \\ \min(1, 2r) & 0 \leq r \leq 1; \\ \min(1.75, r) & 1 < r, \end{cases} \tag{9.4}$$

is more robust and yields results that are similar to those obtained with Superbee (see Sect. 12). Both functions are shown in Fig. 4.

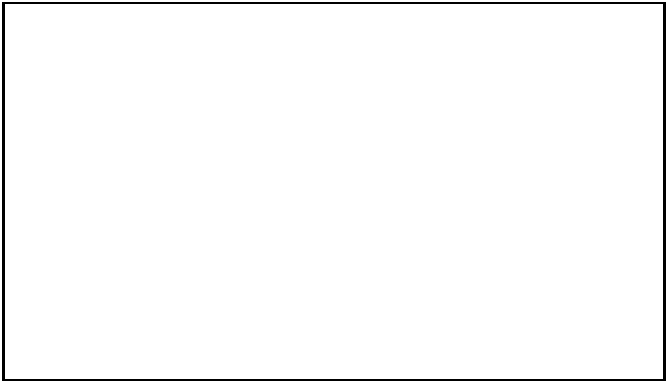

**Fig. 4.** Flux limiter "Superbee". The original and modified "Superbee" interpolation functions are shown as a function of $r$, the ratio of the difference at upwind and the current interface. If the flux limiter is 1 the data is linearly interpolated between to cells; if it is 0 the upwind data is chosen

This form of the flux limiter is not easy to use numerically because the switching parameter is determined by a division (possibly by zero). Of course the final update is well defined, because of the multiplication by $[a_k]^{n+\frac{1}{2}-}_{i+\frac{1}{2}}$ of the flux limiter. A form of Eq. (9.4) that automatically chooses the correct limit is found by replacing the combination $b\psi(a/b)$ with the equivalent

$$b\psi(a/b) = \max(0, \min(2a, \max(b, \min(a, 1.75b))))$$
$$+ \min(0, \max(2a, \min(b, \max(a, 1.75b)))). \tag{9.5}$$

Another possible definition of $r_k$ is

$$r_k = \frac{[b_k]^{n+\frac{1}{2}-}_{i+\frac{1}{2}-\sigma_k}}{[b_k]^{n+\frac{1}{2}-}_{i+\frac{1}{2}}} = \frac{[\lambda_k a_k]^{n+\frac{1}{2}-}_{i+\frac{1}{2}-\sigma_k}}{[\lambda_k a_k]^{n+\frac{1}{2}-}_{i+\frac{1}{2}}}. \tag{9.6}$$

Indeed the different definitions of $r_k$ generally yield comparable results except near sonic points. There $\lambda_k$ changes sign while $a_k$ does not. Thus the second definition of $r_k$ yields $\psi_k = 0$, i.e. shock-like interpolation. This results in unphysical expansion shocks, which allow the entropy to decrease along a streamline. It is well known that expansion shocks are a problem in first order methods of the type described above. Several suggestions for a remedy have appeared in literature (e.g. Roe 1985; Van Leer et al. 1989; Roe 1992). If the first definition is used, linear interpolation is automatically chosen near sonic points and smooth transsonic flow is possible and does in fact occur in numerical experiments. Thus the first definition of $r_k$ is to be preferred.

## 10. Roe solver for the equations of general relativistic fluid dynamics

The method outlined in Sect. 4 can be used to develop a Roe solver for GRH. The vector of all entries of the metric takes the place of the known functions $g$. The vector $F^0$ takes the place of the state vector $W$.

To derive the appropriate matrix $\tilde{\mathbf{A}}$ it is convenient to use a parameter vector $w$ as in the CNR case (Roe 1981b):

$$w \equiv (w^\alpha, w^4) \equiv (w^0, w^1, w^2, w^3, w^4), \tag{10.1}$$

where

$$w^\alpha \equiv K u^\alpha, \qquad w^4 \equiv K \frac{p}{\rho h}, \tag{10.2}$$

and

$$K^2 \equiv \sqrt{-g}\rho h = -g_{\alpha\beta} w^\alpha w^\beta. \tag{10.3}$$

The flow variables can then be expressed in terms of this $w$ by

$$F^\alpha = \left( \left( K - \frac{\Gamma}{\Gamma - 1} w^4 \right) w^\alpha, w^\alpha w^\beta + K w^4 g^{\alpha\beta} \right). \tag{10.4}$$

By repeatedly using Eq. (4.16) one can write for two neighbouring cells

$$F_R^\alpha - F_L^\alpha \equiv \Delta F^\alpha = \mathbf{A}^\alpha \Delta w = \mathbf{A}^\alpha (w_R - w_L). \tag{10.5}$$

The Roe matrix of the first spatial coordinate is

$$\tilde{\mathbf{A}} = \mathbf{A}^1 (\mathbf{A}^0)^{-1}. \tag{10.6}$$

The matrices of the other coordinates follow from symmetry. Notice that Eq. (10.6) implicitly assumes that both parameter vectors are at the same point in spacetime, because terms containing metric *differences* are not allowed for.

Using the new symbols

$$v \equiv \frac{\langle w \rangle}{\langle K \rangle},$$

$$v_\alpha \equiv g_{\alpha\beta} v^\beta,$$

$$c_- \equiv 1 - \frac{\Gamma}{\Gamma - 1} v^4,$$

$$c_+ \equiv 1 + \frac{\Gamma}{\Gamma - 1} v^4, \tag{10.7}$$

the **A**-matrices can be written as follows:

$$\mathbf{A}^0 = \langle K \rangle \begin{pmatrix} c_- - v^0 v_0 & -v^0 v_1 \\ 2v^0 - v^4 g^{00} v_0 & -v^4 g^{00} v_1 \\ v^1 - v^4 g^{01} v_0 & v^0 - v^4 g^{01} v_1 \\ v^2 - v^4 g^{02} v_0 & -v^4 g^{02} v_1 \\ v^3 - v^4 g^{03} v_0 & -v^4 g^{03} v_1 \end{pmatrix}$$

$$\begin{pmatrix} -v^0 v_2 & -v^0 v_3 & -\frac{\Gamma}{\Gamma-1} v^0 \\ -v^4 g^{00} v_2 & -v^4 g^{00} v_3 & g^{00} \\ -v^4 g^{01} v_2 & -v^4 g^{01} v_3 & g^{01} \\ v^0 - v^4 g^{02} v_2 & -v^4 g^{02} v_3 & g^{02} \\ -v^4 g^{03} v_2 & v^0 - v^4 g^{03} v_3 & g^{03} \end{pmatrix}, \quad (10.8)$$

and

$$\mathbf{A}^1 = \langle K \rangle \begin{pmatrix} -v^1 v_0 & c_- - v^1 v_1 \\ v^1 - v^4 g^{01} v_0 & v^0 - v^4 g^{01} v_1 \\ -v^4 g^{11} v_0 & 2v^1 - v^4 g^{11} v_1 \\ -v^4 g^{12} v_0 & v^2 - v^4 g^{12} v_1 \\ -v^4 g^{13} v_0 & v^3 - v^4 g^{03} v_1 \end{pmatrix}$$

$$\begin{pmatrix} -v^1 v_2 & -v^1 v_3 & -\frac{\Gamma}{\Gamma-1} v^1 \\ -v^4 g^{01} v_2 & -v^4 g^{01} v_3 & g^{01} \\ -v^4 g^{11} v_2 & -v^4 g^{11} v_3 & g^{11} \\ v^1 - v^4 g^{12} v_2 & -v^4 g^{12} v_3 & g^{12} \\ -v^4 g^{13} v_2 & v^1 - v^4 g^{13} v_3 & g^{13} \end{pmatrix}, \quad (10.9)$$

and similar expressions for $\mathbf{A}^2$ and $\mathbf{A}^3$.

For determining the right eigenvalues $\lambda_k$ and eigenvectors $e_k$ of the matrix $\tilde{\mathbf{A}}$ one can use the relation

$$(\mathbf{A}^1 - \lambda_k \mathbf{A}^0)(\mathbf{A}^0)^{-1} e_k = 0. \quad (10.10)$$

The eigenvalues are then found to be

$$\lambda_1 = \lambda^-,$$
$$\lambda_2 = \lambda^+,$$
$$\lambda_3 = \lambda_4 = \lambda_5 = \lambda^0, \quad (10.11)$$

where

$$\lambda^\pm \equiv \frac{(1 - \Gamma v^4) v^0 v^1 - s^2 g^{01} \pm s y}{(1 - \Gamma v^4) v^0 v^0 - s^2 g^{00}},$$
$$\lambda^0 \equiv \frac{v^1}{v^0}, \quad (10.12)$$

with

$$s^2 \equiv \tfrac{1}{2} \Gamma v^4 (1 - v^\alpha v_\alpha) - \tfrac{1}{2}(\Gamma - 1)(1 + v^\alpha v_\alpha), \quad (10.13)$$

and

$$y^2 \equiv (1 - \Gamma v^4) e + s^2 (g^{01} g^{01} - g^{00} g^{11}),$$
$$e \equiv g^{00} v^1 v^1 - 2 g^{01} v^0 v^1 + g^{11} v^0 v^0. \quad (10.14)$$

The quantity $s$ is the sound speed in the comoving inertial frame. Notice that since all these quantities are averages of two determinations, in general $v^\alpha v_\alpha \neq -1$! The corresponding eigenvectors are

$$e_1 = \left( c_-, v^\alpha - \frac{s}{y}(g^{1\alpha} v^0 - g^{0\alpha} v^1) \right)$$

$$e_2 = \left( c_-, v^\alpha + \frac{s}{y}(g^{1\alpha} v^0 - g^{0\alpha} v^1) \right)$$

$$e_3 = \left( c_- + \frac{s^2}{\Gamma - 1}, v^\alpha \right)$$

$$e_4 = \left( -c_+ v_2, 0, 0, 1, 0 \right)$$

$$e_5 = \left( -c_+ v_3, 0, 0, 0, 1 \right). \quad (10.15)$$

The decomposition of a vector $d \equiv (\delta, \Delta^0, \Delta^1, \Delta^2, \Delta^3)$ in terms of these eigenvectors is given by $d = \sum a_k e_k$, where

$$a_1 = \frac{-1}{2es^2} \left( s^2 k + sy(v^0 \Delta^1 - v^1 \Delta^0) + (\Gamma - 1) e(\delta + c_+ v_\alpha \Delta^\alpha) \right)$$

$$a_2 = \frac{-1}{2es^2} \left( s^2 k - sy(v^0 \Delta^1 - v^1 \Delta^0) + (\Gamma - 1) e(\delta + c_+ v_\alpha \Delta^\alpha) \right)$$

$$a_3 = \frac{1}{es^2} \left( 2s^2 k + (\Gamma - 1) e(\delta + c_+ v_\alpha \Delta^\alpha) \right)$$

$$a_4 = \Delta^2 + \frac{1}{e} \left( (g^{02} v^1 - g^{12} v^0)(v^0 \Delta^1 - v^1 \Delta^0) - k v^2 \right)$$

$$a_5 = \Delta^3 + \frac{1}{e} \left( (g^{03} v^1 - g^{13} v^0)(v^0 \Delta^1 - v^1 \Delta^0) - k v^3 \right), \quad (10.16)$$

$$k \equiv g^{00} v^1 \Delta^1 - g^{01}(v^0 \Delta^1 + v^1 \Delta^0) + g^{11} v^0 \Delta^0. \quad (10.17)$$

The numerical ease of this decomposition is related to the determinant of the set of eigenvectors. This determinant is $-2es^3/[(\Gamma - 1)y]$. Hence decomposition is numerically most difficult for low Mach numbers.

To decompose the state difference one should take

$$d = [F^0]_{i+\frac{1}{2}+}^{n+\frac{1}{2}-} - [F^0]_{i+\frac{1}{2}-}^{n+\frac{1}{2}-}. \quad (10.18)$$

This completes the GR Roe solver. The formulae that are needed for application rather than derivation are: (10.15), (10.16) and (10.18) to find the decomposition of the state difference (or equivalently the flux difference) along the characteristics.

## 11. Limiting cases

Before discussing test cases, it is useful and straightforward to deduce the Roe solver for a number of important limiting cases. It shown in Appendix E that in the limit of smooth flow the characteristic velocities found by the Roe solver are indeed the characteristics of one-dimensional GRH. Below the Roe solver for a number of special situations is derived by considering the appropriate limit of the GR Roe solver.

## 11.1. Special relativistic Roe solver

To derive a Roe solver for the equations of special relativistic fluid dynamics, all one needs to do is to replace the general metric $g^{\alpha\beta}$ by the Minkowski metric $\eta^{\alpha\beta}$:

$$g^{\alpha\beta} = g_{\alpha\beta} \equiv \eta^{\alpha\beta},$$
$$\sqrt{-g} = 1,$$
$$\Gamma^{\alpha}_{\beta\gamma} = 0, \tag{11.1}$$

This yields

$$v_0 = -v^0,$$
$$v_j = v^j,$$
$$e = v^0 v^0 - v^1 v^1,$$
$$k = v^0 \Delta^0 - v^1 \Delta^1. \tag{11.2}$$

Apart from the substitution of the metric, the expressions for the eigenvectors and the coefficients cannot be simplified further. The expressions thus obtained are similar to the ones discussed in Martí et al. 1991 and Marquina et al. 1992.

## 11.2. Non relativistic Roe solver in arbitrary spatial coordinates

In order to obtain the non relativistic limit, consider a weak gravitational field $\Phi \ll 1$, in a Newtonian coordinate frame ($X^{\alpha}$). The metric is (Misner et al. 1973, 18.4):

$$-(1 + 2\Phi)dX^0 dX^0 + (1 - 2\Phi)\delta_{ij}dX^i dX^j, \tag{11.3}$$

where $\Phi$ is the gravitational potential normalized to zero at infinity.

The gravitational potential in the spatial part of the metric does not influence the motion of non relativistic fluids because it and its contribution to the Christoffel symbols enter only in terms that may be neglected in the non relativistic limit. Thus, for a frame with arbitrary spatial coordinates and Newtonian time ($x^0 = X^0$, $x^i$) one has

$$g_{00} = -1 - 2\Phi\delta_{kl}\frac{\partial X^k}{\partial x^0}\frac{\partial X^l}{\partial x^0},$$
$$g_{0i} = \frac{\partial X^k}{\partial x^i}\frac{\partial X^l}{\partial x^0},$$
$$g_{ij} = \frac{\partial X^k}{\partial x^i}\frac{\partial X^l}{\partial x^j}. \tag{11.4}$$

The spatial part of the four-metric is the three-metric that leaves distance invariant. If this is a frame that moves non relativistically with respect to a Newtonian frame, i.e. $g_{0i} \ll \sqrt{g_{ii}}$, one obtains $-g \to det(g_{ij}) = g_c$.

One arrives at the non relativistic limit if in these coordinates both the mean and the random velocities of the particles are small compared to the speed of light:

$$\frac{p}{\rho} \ll 1,$$
$$g_{ij} u^i u^j \ll 1. \tag{11.5}$$

As a result,

$$K^2 \to \sqrt{g_c}\rho \equiv K_c^2. \tag{11.6}$$

Define the non relativistic enthalpy by

$$h_c \equiv \tfrac{1}{2}\left[g_{ij}u^i u^j + (1 + g_{00})\right] + g_{0i}u^i + \frac{\Gamma}{\Gamma - 1}\frac{p}{\rho}. \tag{11.7}$$

Then

$$u^0 \to 1 + h_c - \frac{\Gamma}{\Gamma - 1}\frac{p}{\rho},$$
$$w^j \to K_c u^j \equiv w_c^j,$$
$$w^4 \to K_c \frac{p}{\rho} \equiv w_c^4,$$
$$T^{00} \to \rho,$$
$$T^{0i} = T^{i0} \to \rho u^i,$$
$$T^{ij} \to \rho u^i u^j + p g^{ij},$$
$$v^0 \to 1 + \frac{\langle K_c h_c \rangle}{\langle K_c \rangle} - \frac{\Gamma}{\Gamma - 1}v_c^4 \equiv 1 + \tilde{h}_c - \frac{\Gamma}{\Gamma - 1}v_c^4,$$
$$v^j \to \frac{\langle w_c^j \rangle}{\langle K_c \rangle} \equiv v_c^j,$$
$$v^4 \to \frac{\langle w_c^4 \rangle}{\langle K_c \rangle} \equiv v_c^4,$$
$$c_- \to 1,$$
$$c_+ \to 1,$$
$$y^2 \to e \to g^{11}. \tag{11.8}$$

The speed of sound is

$$s^2 \to (\Gamma - 1)\left\{\tilde{h}_c - \left[g_{0i}v_c^i + \tfrac{1}{2}(g_{ij}v_c^i v_c^j + 1 + g_{00})\right]\right\} \equiv s_c^2. \tag{11.9}$$

Notice that both the enthalpy and the sound speed are three-scalars that depend only on the cell choice through averaging, but not on the three-metric itself. With this sound speed the eigenvalues become

$$\lambda^{\pm} \to v_c^1 \pm \sqrt{g^{11}}s_c,$$
$$\lambda^0 \to v_c^1. \tag{11.10}$$

Substitution of these limits in the eigenvectors yields, to first order, a set of linearly dependent eigenvectors. This reflects the well-known fact that the total energy is dominated by the rest mass energy in the classical limit. Because the latter cannot classically be converted into other forms, it can be subtracted from the total energy:

$$\tilde{T}^{00} = \rho h u^0 u^0 + p g^{00} - \rho u^0 \to \rho h_c - p,$$
$$\tilde{T}^{0i} = \rho h u^0 u^i + p g^{0i} - \rho u^i \to \rho h_c + p g^{0i}. \tag{11.11}$$

This amounts to subtracting the first component of the eigenvectors from the second. The conserved vector quantity is now

$$F^0 = \sqrt{g_c} \left( \rho, \rho h_c - p, \rho u^i \right). \tag{11.12}$$

The flux vector in the first direction is

$$F^1 = \sqrt{g_c} \left( \rho u^1, \rho h_c u^1 + p g^{01}, \rho u^1 u^i + p g^{1i} \right), \tag{11.13}$$

while the NR limit of the source vector is given by

$$S = \sqrt{g_c} \left( 0, S^\alpha - \rho(\Gamma_{00}^\alpha + 2\Gamma_{0i}^\alpha u^i) - \Gamma_{ij}^\alpha T^{ij} \right), \tag{11.14}$$

and the NR limit of the eigenvectors is

$$e_1 \to \left( 1, \tilde{h}_c - \frac{s_c}{\sqrt{g^{11}}}(v_c^1 + g^{01}), v_c^i - \frac{s_c}{\sqrt{g^{11}}} g^{1i} \right),$$
$$e_2 \to \left( 1, \tilde{h}_c + \frac{s_c}{\sqrt{g^{11}}}(v_c^1 + g^{01}), v_c^i + \frac{s_c}{\sqrt{g^{11}}} g^{1i} \right),$$
$$e_3 \to \left( 1, \frac{1}{2}(g_{kl} v_c^k v_c^l + 1 + g_{00}) + g_{0k} v_c^k, v_c^i \right),$$
$$e_4 \to \left( 0, v_{2c}, 0, 1, 0 \right),$$
$$e_5 \to \left( 0, v_{3c}, 0, 0, 1 \right). \tag{11.15}$$

Let $\tilde{\Delta}^0 \equiv \Delta^0 - \delta$ denote the difference in the energy equation after subtraction of the rest mass energy. The coefficients of the projection reduce to

$$a_1 \to \frac{\Gamma - 1}{2 s_c^2} \left\{ \tfrac{1}{2} \left[ g_{ij} v_c^i v_c^j - (1 + g_{00}) \right] \delta + \tilde{\Delta}^0 - v_{jc} \Delta^j \right\}$$
$$- \frac{\Delta^1 - v_c^1 \delta}{2 s_c \sqrt{g^{11}}},$$
$$a_2 \to \frac{\Gamma - 1}{2 s_c^2} \left\{ \tfrac{1}{2} \left[ g_{ij} v_c^i v_c^j - (1 + g_{00}) \right] \delta + \tilde{\Delta}^0 - v_{jc} \Delta^j \right\}$$
$$+ \frac{\Delta^1 - v_c^1 \delta}{2 s_c \sqrt{g^{11}}},$$
$$a_3 \to \frac{\Gamma - 1}{s_c^2} \left[ (\tilde{h}_c - g_{0i} v_c^i - g_{ij} v_c^i v_c^j) \delta + v_{jc} \Delta^j - \tilde{\Delta}^0 \right],$$
$$a_4 \to \Delta^2 - v_c^2 \delta - \frac{g^{12}}{g^{11}}(\Delta^1 - v_c^1 \delta),$$
$$a_5 \to \Delta^3 - v_c^3 \delta - \frac{g^{13}}{g^{11}}(\Delta^1 - v_c^1 \delta). \tag{11.16}$$

Three further simplifications of these formulae are possible. One can consider a fluid with no background gravitational field by choosing $\Phi = 0$ and/or a fluid with no inertial accelerations by taking $g_{0i} = 0$ and/or look at the equations in a Cartesian frame. These limits are easily obtained by inserting the appropriate metric. The expressions for the case of spherical coordinates and no gravity are used in Mellema et al. (1991). The simplest case, i.e., the combination of all three simplifications, is CNRH. In that limit the expressions are exactly those found by Roe (1981b), which were discussed in Sect. 4. Notice that this is more than just a verification that the Cartesian limit of the formulae is correct: the correspondence also hinges on the choice of parameter vectors. These were chosen in such a way that the non relativistic Cartesian limit is a linear combination of the elements of the parameter vector used by Roe:

$$w \to K_C \left( 1 + h_c - \frac{\Gamma}{\Gamma - 1} \frac{p}{\rho}, v_c^1, v_c^2, v_c^3, \frac{p}{\rho} \right). \tag{11.17}$$

### 11.3. Motion in a spherically symmetric gravitational potential

To describe spherically symmetric motion of a fluid under the influence of a spherically symmetric gravitational potential, it is convenient to choose spherical coordinates ($x^0 = t, x^1 = r, x^2 = \theta, x^3 = \phi$) that are related to Cartesian coordinates ($X^\alpha$) by

$$X^0 = t,$$
$$X^1 = r \sin\theta \cos\phi,$$
$$X^2 = r \sin\theta \sin\phi,$$
$$X^3 = r \cos\theta. \tag{11.18}$$

The only non-zero elements of the covariant metric tensor are

$$g_{00} = -1 - 2\Phi,$$
$$g_{11} = 1,$$
$$g_{22} = r^2,$$
$$g_{33} = r^2 \sin^2\theta, \tag{11.19}$$

and the volume element is $\sqrt{g_c} = r^2 \sin\theta$. Note that in fact the non relativistic limit has $g_{11} = 1 - 2\Phi$ but the gravitational potential and the extra non-zero Christoffel symbols it induces are too small to make a contribution to the NR limit. The non-zero Christoffel symbols are

$$\Gamma^0_{00} = \frac{\partial \Phi}{\partial t},$$

$$\Gamma^0_{01} = \Gamma^0_{10} = \frac{\partial \Phi}{\partial r},$$

$$\Gamma^1_{00} = \frac{\partial \Phi}{\partial r},$$

$$\Gamma^1_{22} = -r,$$

$$\Gamma^1_{33} = -r \sin^2 \theta,$$

$$\Gamma^2_{12} = \Gamma^2_{21} = \frac{1}{r},$$

$$\Gamma^2_{33} = -\sin \theta \cos \theta,$$

$$\Gamma^3_{13} = \Gamma^3_{31} = \frac{1}{r},$$

$$\Gamma^3_{23} = \Gamma^3_{32} = \cot \theta. \tag{11.20}$$

A spherically symmetric solution to these equations is found if the source vector and the initial- and boundary conditions are spherically symmetric ($u^2 = u^3 = 0$). Then the only non-zero components of the stress-energy tensor are

$$\tilde{T}^{00} = \rho h_c - p,$$

$$\tilde{T}^{01} = \rho h_c u^1,$$

$$T^{10} = \rho u^1,$$

$$T^{11} = \rho u^1 u^1 + p,$$

$$T^{22} = \frac{p}{r^2},$$

$$T^{33} = \frac{p}{r^2 \sin^2 \theta}. \tag{11.21}$$

This yields, after dividing out $\sin \theta$, the non-trivial vectors

$$F^0 = r^2 \left( \rho, \rho h_c - p, \rho u^1 \right),$$
$$F^1 = r^2 \left( \rho u^1, \rho h_c u^1, \rho u^1 u^1 + p \right), \tag{11.22}$$

and the source term according to Eq. (11.14)

$$S = r^2 \left( 0, S^0 - \rho \frac{\partial \Phi}{\partial t} - 2\rho u^1 \frac{\partial \Phi}{\partial r}, S^1 + 2\frac{p}{r} - \rho \frac{\partial \Phi}{\partial r} \right). \tag{11.23}$$

The quantities for numerical approximation become

$$K_c^2 = r^2 \rho,$$

$$h_c = -\Phi + \tfrac{1}{2} u^1 u^1 + \frac{\Gamma}{\Gamma - 1} \frac{p}{\rho},$$

$$w^1 = K_c u^1,$$

$$w^4 = K_c \frac{p}{\rho},$$

$$s_c^2 = (\Gamma - 1) \left( \tilde{h}_c - \tfrac{1}{2} v_c^1 v_c^1 + \tilde{\Phi} \right). \tag{11.24}$$

The characteristic speeds are

$$\lambda_1 = v_c^1 - s_c,$$
$$\lambda_2 = v_c^1 + s_c,$$
$$\lambda_3 = v_c^1, \tag{11.25}$$

corresponding to the eigenvectors

$$e_1 = \left( 1, \tilde{h}_c - s_c v_c^1, v_c^1 - s_c \right),$$
$$e_2 = \left( 1, \tilde{h}_c + s_c v_c^1, v_c^1 + s_c \right),$$
$$e_3 = \left( 1, \tfrac{1}{2} v_c^1 v_c^1 - \tilde{\Phi}, v_c^1 \right). \tag{11.26}$$

The projection coefficients are then given by

$$a_1 = \frac{\Gamma - 1}{2 s_c^2} \left( [\tfrac{1}{2} v_c^1 v_c^1 + \Phi] \delta + \tilde{\Delta}^0 - v_c^1 \Delta^1 \right) - \frac{\Delta^1 - v_c^1 \delta}{2 s_c},$$

$$a_2 = \frac{\Gamma - 1}{2 s_c^2} \left( [\tfrac{1}{2} v_c^1 v_c^1 + \Phi] \delta + \tilde{\Delta}^0 - v_c^1 \Delta^1 \right) + \frac{\Delta^1 - v_c^1 \delta}{2 s_c},$$

$$a_3 = \frac{\Gamma - 1}{s_c^2} \left( [\tilde{h}_c - v_c^1 v_c^1] \delta + v_c^1 \Delta^1 - \tilde{\Delta}^0 \right). \tag{11.27}$$

## 12. Test problems

In this section the performance of the general relativistic Roe solver is tested on problems to which an exact solution is known. These test problems are listed in HSW1 and were used by these authors to test five numerical schemes (HSW2). For the precise description of all test problems and their analytical solutions we refer to these references. The results of our GR Roe solver can thus be compared to both the analytical solution of HSW1 and the numerical results of HSW2. HSW usually quote the relative errors after averaging over the relevant region. These errors are not sensitive to zone to zone oscillations. Therefore we present the maximum relative errors. With either error definition regions where the analytical value of a quantity is zero have to be excluded. A measure of the absolute errors there may be obtained from the figures.

Notice that both the test problems and the analytical solutions themselves are dissipation free. The artificial viscosity that explicitly enters some numerical schemes of HSW is only a numerical tool to damp unwanted oscillations near shocks. Although all numerical schemes have a certain degree of numerical viscosity, no explicitly introduced viscosity is needed to obtain almost oscillation free shocks in the present scheme. In all test problems there is no external source term ($S^\alpha = 0$) and no time dependence of the metric. The *same scheme* based on the GR Roe solver was used on all test problems. Hence, the GR algebraic extrapolation routines, for instance, were called even in test problems that could be solved within the framework of CNRH. Such problems thus provide an essential test of the NR limit of the GR scheme.

**Table 1.** The maximal relative errors of the Roe solver and several HSW schemes for the Riemann shock tube (test problem 1). HSW errors have been estimated from their figures

| method | $\Delta t$ | $\Delta x^1$ | $\rho \frac{u^0}{\sqrt{-g^{00}}}$ | $\frac{u^1}{u^0}$ | $e$ | $p$ | $\rho h u_1 \frac{u^0}{\sqrt{-g^{00}}}$ |
|---|---|---|---|---|---|---|---|
| Roe | 25 | 1 | 0.75% | 0.21% | 0.75% | 0.27% | 0.76% |
| HSW Wilson | 25 | 1 | 7% | 8% | 7% | 7% | 14% |
| HSW Barton | 25 | 1 | 2% | 2% | 3% | 2% | 2% |
| HSW Mono | 25 | 1 | 4% | 3% | 2% | 4% | 2% |
| Roe | 12.5 | 0.5 | 0.58% | 0.18% | 0.58% | 0.24% | 0.58% |
| HSW Wilson | 12.5 | 0.5 | 7% | 8% | 7% | 7% | 11% |
| HSW Barton | 12.5 | 0.5 | 2% | 2% | 3% | 2% | 1% |
| HSW Mono | 12.5 | 0.5 | 3% | 3% | 2% | 3% | 2% |

**Table 2.** The maximal relative errors of the Roe solver and several HSW schemes for the Riemann strong shock tube (test problem 2). HSW errors have been estimated from their figures

| method | $\Delta t$ | $\Delta x^1$ | $\rho \frac{u^0}{\sqrt{-g^{00}}}$ | $\frac{u^1}{u^0}$ | $e$ | $p$ | $\rho h u_1 \frac{u^0}{\sqrt{-g^{00}}}$ |
|---|---|---|---|---|---|---|---|
| Roe | 0.75 | 1 | 6.62% | 1.45% | 8.60% | 3.19% | 6.01% |
| HSW Mono | 0.75 | 1 | 12% | 5% | 21% | 25% | 17% |
| Roe | 0.15 | 0.2 | 7.24% | 0.28% | 6.72% | 1.12% | 1.69% |
| HSW Mono | 0.15 | 0.2 | 6% | 2% | 10% | 12% | 11% |

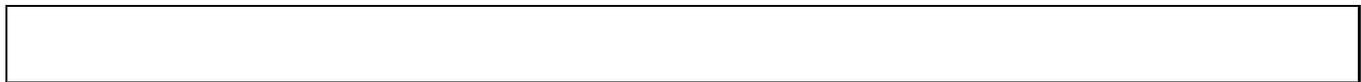

**Fig. 5.** Comparison of the numerical (crosses) and analytical solution of the Riemann shock problem on a 100 zone grid (test problem 1)

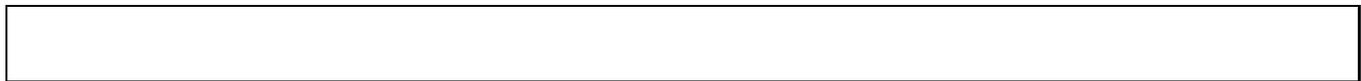

**Fig. 6.** Comparison of the numerical (crosses) and analytical solution of the Riemann shock problem on a 200 zone grid (test problem 1)

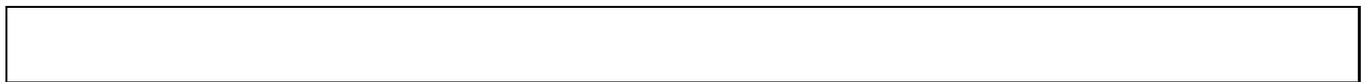

**Fig. 7.** Comparison of the numerical (crosses) and analytical solution of the Riemann strong shock problem on a 100 zone grid (test problem 2)

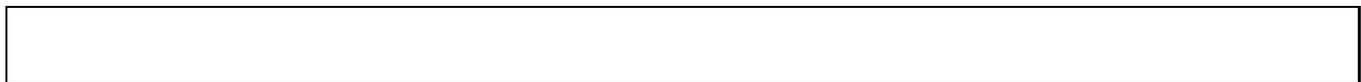

**Fig. 8.** Comparison of the numerical (crosses) and analytical solution of the Riemann strong shock problem on a 500 zone grid (test problem 2)

### 12.1. Non relativistic Riemann shock tube tests

Test problem 1 is a non relativistic shock in a gas with $\Gamma = 7/5$. The physical domain stretches between $x^1 = 0$ and $x^1 = 100$.

Across a membrane at $x^1 = 35$, which is removed at $x^0 = 0$, the gas is characterized by

$$(\rho u^0)_L = 1\ 10^5, \qquad (\rho u^0)_R = 1.25\ 10^4,$$
$$u_L^1 = 0, \qquad u_R^1 = 0, \qquad (12.1)$$
$$p_L = 1, \qquad p_R = 0.1.$$

The numerical solution is calculated on a 100 zone grid with $\Delta x^1 = 1$ and a time step of $\Delta x^0 = 25$. The solution at $x^0 = 5000$ is compared to the analytical solution in Fig. 5.

The contact discontinuity is spread over 4 cells instead of the 6 cells in their best solution and at the same time the zone to zone oscillations are far less pronounced. The solution is already so accurate that little improvement may be expected from further grid refinement. For the sake of comparison the above test problem is also solved on a 200 zone $\Delta x^1 = 0.5$ grid with half the time step used above: $\Delta x^0 = 12.5$. The solution, again at $x^0 = 5000$, is plotted in Fig. 6. Compared to the solution on the 100 zone grid there are indeed hardly any differences. The errors are shown in Table 1.

The Roe solver produces solutions that are almost an order of magnitude more accurate than the best solutions of HSW. The solutions of all schemes improve relatively little on a finer grid. This is an indication that the errors are dominated by the discontinuities, which are scale free.

Test problem 2 considers a strong non relativistic shock problem in a gas with $\Gamma = 5/3$. It is specified by:

$$(\rho u^0)_L = 1\ 10^2, \qquad (\rho u^0)_R = 1,$$
$$u_L^1 = 0, \qquad u_R^1 = 0, \qquad (12.2)$$
$$p_L = 2/3, \qquad p_R = 2/3\ 10^{-7}.$$

The numerical solution is calculated on a 100 zone grid with $\Delta x^1 = 1$ and a time step of $\Delta x^0 = 0.75$. In this problem the method gives a negative pressure in cell 38 after 4 time steps if the flux limiter as suggested by Roe is used. A negative pressure is lethal to a characteristic-based scheme since the speed of sound cannot be determined. By using the slightly modified form of Sect. 9 no problems arise. If one applies the original Superbee on the coefficients of the flux difference $(\lambda_k a_k)$ also no negative pressures arise, but with this modification expansion shocks form instead of transsonic expansion waves. The solution for $x^0 = 172.5$ is compared to the analytical solution in Fig. 7. Results on a 500 zone grid with $\Delta x^0 = 0.15$, are shown in Fig. 8.

The performance on this test problem shows the ability of the relativistic Roe scheme to accurately form and propagate NR shocks, contact discontinuities and rarefaction waves. The errors are given in Table 2. Again the Roe solver compares favourably to the HSW scheme. This time the contact discontinuity and the shock are hardly separated on the coarse grid and consequently the accuracy is improved on the finer grid. Nevertheless the results of the Roe solver on a 100 zone grid are comparable to (if not better than) the HSW solution on a 500 zone grid.

*12.2. Special relativistic Riemann shock tube tests*

Next, four special relativistic problems were treated. A relativistically moving gas with $\Gamma = 5/3$ collides at the centre of the grid ($x^1 = 50$) at $x^0 = 0$ with a similar gas moving at an equal speed in the opposite direction. This problem is completely equivalent to the wall shock problem of HSW, except that it is presented in a boundary independent way. The initial condition is

$$(\rho u^0)_L = 1, \qquad (\rho u^0)_R = 1,$$
$$u_L^1 = \sqrt{u^0 u^0 - 1}, \qquad u_R^1 = -\sqrt{u^0 u^0 - 1}, \qquad (12.3)$$
$$p_L = 2/3\ 10^{-6}, \qquad p_R = 2/3\ 10^{-6}.$$

First, we take $u^0 = 2.24$. The numerical scheme was run on a 100 zone $\Delta x^1 = 1$ grid with a time step of $\Delta x^0 = 0.5$. The result is shown in Fig. 9. The most prominent error is the spike in internal energy at the point of collision. This "wall heating" phenomenon is the actual solution of the finite difference equations as was shown by Norman & Winkler (1986), and appears in all methods tested by HSW as well. Apart from this spike, the result is excellent compared to the analytical solution. Contrary to the solutions of HSW the post shock density and specific internal energy do not have a significant systematic error. In addition the zone to zone oscillations are far less pronounced than with the HSW scheme. The errors are given in Table 3.

The Roe errors are typically a factor three less than those of HSW. The methods of HSW become unacceptably unstable around $u^0 = 4$. They reach the same conclusion as Norman & Winkler (1986, presented in 1983), namely that extremely relativistic shocks could -at that time- only be reproduced by implicit codes, which are much more expensive than explicit methods. The present explicit method, however, performs well on extremely relativistic shocks: results for $u^0 = 625$ and $\Gamma = 5/3$ are shown in Fig. 10 and for $u^0 = 625$ and $\Gamma = 4/3$ in Fig. 11.

The errors are given in Table 4 and are in fact somewhat smaller than those at $u^0 = 2.24$. This is caused by the fact that the Courant number of these calculations is closer to unity, which has a favourable effect.

Recently, Marquina et al. (1992) have solved this test problem up to $u^0 = 70$ with a somewhat different implementation of a Roe solver. Hence, the Roe method is to our knowledge the first explicit scheme that can deal with extremely relativistic shocks. This is strong evidence that the advantages of the characteristic based schemes for violent NR flows (Woodward & Colella 1984) are retained by the relativistic version of the Roe solver.

In test problem 4 a special relativistic shock in a gas with $\Gamma = 5/3$ is considered. It is specified by a membrane at $x^1 = 45$ and:

$$(\rho u^0)_L = 10, \qquad (\rho u^0)_R = 1,$$
$$u_L^1 = 0, \qquad u_R^1 = 0, \qquad (12.4)$$
$$p_L = 40/3, \qquad p_R = 2/3\ 10^{-7}.$$

The numerical solution is calculated on a 100 zone grid with $\Delta x^1 = 1$ and a time step of $\Delta x^0 = 0.5$. The solution is for

**Table 3.** The maximal relative errors of the Roe solver and several HSW schemes for the mildly relativistic ($u^0 = 2.24$) collision (test problem 3a). HSW errors have been estimated from their figures

| method | $\Delta t$ | $\Delta x^1$ | $\rho \frac{u^0}{\sqrt{-g^{00}}}$ | $\frac{u^1}{u^0}$ | $e$ | $p$ | $\rho h u_1 \frac{u^0}{\sqrt{-g^{00}}}$ |
|---|---|---|---|---|---|---|---|
| Roe | 0.5 | 1 | 8.72% | 0.00% | 9.58% | 0.15% | 0.00% |
| HSW Mono | 0.5 | 1 | 20% | 0% | 22% | 3% | 4% |

**Table 4.** The maximal relative errors of the Roe solver and several HSW schemes for the ultra relativistic ($u^0 = 625$) collision (test problems 3b and 3c). HSW schemes, applied to this problem, are unacceptably unstable

| method | $\Delta t$ | $\Delta x^1$ | $\rho \frac{u^0}{\sqrt{-g^{00}}}$ | $\frac{u^1}{u^0}$ | $e$ | $p$ | $\rho h u_1 \frac{u^0}{\sqrt{-g^{00}}}$ |
|---|---|---|---|---|---|---|---|
| Roe ($\Gamma = 5/3$) | 0.5 | 1 | 4.16% | 0.00% | 4.34% | 0.01% | 0.00% |
| Roe ($\Gamma = 4/3$) | 0.5 | 1 | 4.99% | 0.00% | 5.25% | 0.21% | 0.00% |

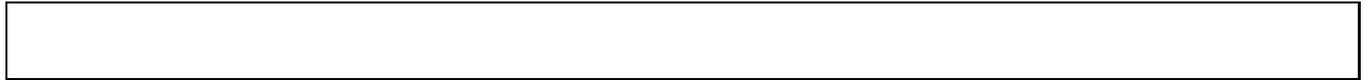

**Fig. 9.** Comparison of the numerical (crosses) and analytical solution of the mildly relativistic ($u^0 = 2.24$) collision (test problem 3a) on a 100 zone grid

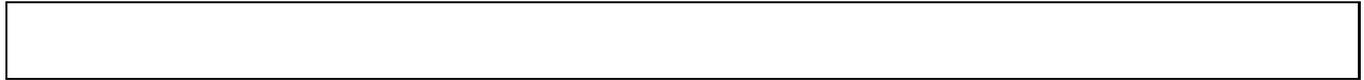

**Fig. 10.** Comparison of the numerical (crosses) and analytical solution of the ultra relativistic ($u^0 = 625$) collision with $\Gamma = 5/3$ (test problem 3b) on a 100 zone grid

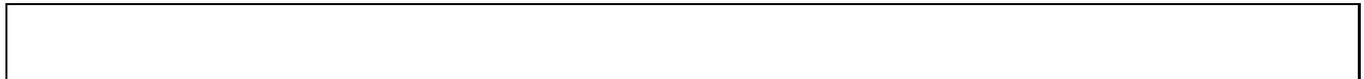

**Fig. 11.** Comparison of the numerical (crosses) and analytical solution of the ultra relativistic ($u^0 = 625$) collision with $\Gamma = 4/3$ (test problem 3c) on a 100 zone grid

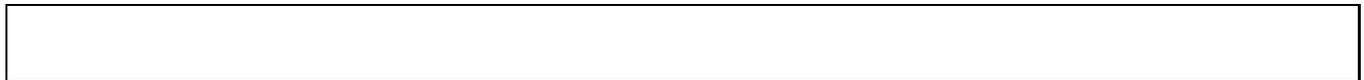

**Fig. 12.** Comparison of the numerical (crosses) and analytical solution of the relativistic shock tube (test problem 4) on a 100 zone grid

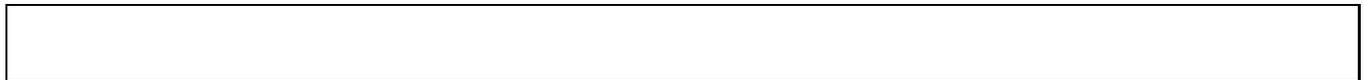

**Fig. 13.** Comparison of the numerical (crosses) and analytical solution of the relativistic shock tube (test problem 4) on a 500 zone grid

$x^0 = 50$ is compared to the analytical solution in Fig. 12. The solution on a 500 zone grid with $\Delta x^1 = 0.2$ and a time step of $\Delta x^0 = 0.1$ is compared to the analytical solution in Fig. 13. The errors in Table 5 show that the Roe solver results on the 100 zone grid are better than the HSW results on a 500 zone grid. The improvement in the Roe results on the finer grid is again due to the fact that the contact discontinuity and the shock were not well separated on the coarse mesh.

**Table 5.** The maximal relative errors of the Roe solver and several HSW schemes for the relativistic shock tube (test problem 4). No HSW results are available on a 100 zone grid. HSW errors have been estimated from their figures

| method | $\Delta t$ | $\Delta x^1$ | $\rho \frac{u^0}{\sqrt{-g^{00}}}$ | $\frac{u^1}{u^0}$ | $e$ | $p$ | $\rho h u_1 \frac{u^0}{\sqrt{-g^{00}}}$ |
|---|---|---|---|---|---|---|---|
| Roe | 0.5 | 1 | 3.00% | 0.69% | 1.51% | 2.71% | 1.94% |
| Roe | 0.1 | 0.2 | 0.04% | 0.66% | 1.05% | 2.58% | 0.11% |
| HSW Mono | 0.1 | 0.2 | 7% | 3% | 14% | 8% | 13% |

*12.3. General relativistic tests*

In test problems 5 and 6 the performance of the code in a general relativistic setting is tested. Both problem are situated in a Schwarzschild metric. The region of $x^1 = 2$ to $x^1 = 18$ is covered by 16 equidistant cells.

In problem 5 radial geometric infall of dust is treated. The analytical solution is again discussed by HSW. Unfortunately their last expression for the analytic value of the energy (HSW1: Eq. [67]) is incorrect. The correct expressions are, in the present notation ($x^1 = r$):

$$d(x^1) = \frac{\rho u^0}{\sqrt{-g^{00}}} = \frac{D}{(x^1)^2 \sqrt{\frac{2}{x^1}\left(1 - \frac{2}{x^1}\right)}},$$

$$e(x^1) = \frac{p}{\Gamma - 1} \frac{u^0}{\sqrt{-g^{00}}} = \frac{\kappa D^\Gamma/(\Gamma-1)}{\left((x^1)^2 \sqrt{\frac{2}{x^1}}\right)^\Gamma \sqrt{\left(1 - \frac{2}{x^1}\right)}}, \quad (12.5)$$

$$v(x^1) = \frac{u^1}{u^0} = \sqrt{\frac{2}{x^1}}\left(1 - \frac{2}{x^1}\right).$$

As an example the test shown by HSW, with boundary condition $\Gamma = 5/3$, $D = 1.6\,10^{-2}$ $h u_0 = -1$ and $\kappa = 12/23$ was run. The initial condition of the gas is given by $D = 1.6\,10^{-4}$ $h u_0 = -1$ and $\kappa = 12/23$. The problem was solved with two different time steps $\Delta x^0 = 0.1875$ and $\Delta x^0 = 0.5$. The result at $x^0 = 90$ is shown in Fig. 14. The errors are summarized in Table 6.

The errors in the dynamically dominant density and velocity are extremely small ($< 0.1\%$). The relative error in the energy is still small, but large compared to the other relative errors (3%). This is due to the fact that in the conservative form of the equations the total energy is solved for. Hence the error in the internal energy is dominated by small relative small errors in the much larger density and velocity. In the (non conservative) formulation of HSW the rest mass energy is subtracted from the total energy and then the internal energy is solved for. This is numerically more benign in case of a dynamically insignificant pressure. The errors on the coarser grid are less than on the finer grid. This is explained by the fact that the fluid is to evolve to a steady state, which analytically is only reached after an infinite amount of time, due to time dilatation near the black hole horizon. Numerical diffusion, which is larger on the coarse grid, accelerates this evolution. This is not the explanation for the HSW errors, because they have reached a numerical steady state at the time the calculation is stopped. If given sufficient time the errors of the present scheme become arbitrarily small by virtue of the algebraic stationary extrapolation subgrid model.

In this problem the use of algebraic extrapolation really pays off: implementations based on numerical integration of the source term (see Appendix C) succeed in reproducing the dynamically dominant density and velocity but fail to reproduce the (dynamically insignificant) pressure, due to numerical diffusion.

To test the ability of the present method to deal with transsonic accretion onto a black hole, a Bondi-Hoyle accretion problem was solved (test problem 6). This differs from geometric infall in the sense that the pressure is no longer negligible. The pressure support results in a lower accretion velocity. The analytical solution is found by assuming a radius for the sonic point $x^1_s$, a value for the radial component of the four-velocity $u^1_s$ and for the entropy constant $\kappa$. Given the critical radius, the metric at the sonic point is known: $g^{00} = -1/(1 - 2/x^1_s)$, $g^{11} = 1 - 2/x^1_s$. Upon substitution of the metric and the assumed $u^1_s$, Eq. (6.23) yields $h/h_0$. The latter constant and the requirement of sonic flow combine in Eq. (6.25) to yield the enthalpy $h$. Using the assumed $\kappa$, the sonic point density $\rho_s$ may be found from Eq. (6.20). The pressure at the sonic point is found from Eq. (6.10). With all these variables known the two remaining constants of the flow ($D$ and $h u_0$) are easily found. The Bondi-Hoyle problem HSW show is given by: $\Gamma = 5/3$, $x^1_s = 8$, $u^1_s = 1/4$ and $\kappa = 120/23$. This yields $h_s = 26/23$, $\rho_s = 1\,10^{-3}$ and flow constants $D \equiv (x^1_s)^2 u^1_s \rho_s = 1.6\,10^{-2}$, $h u_0 \equiv h_s(u_0)_s = -26/23\sqrt{13/16}$. It may be verified that these values are indeed a solution of Eq. (6.11). As an initial condition the analytic solution for $D = 1.6\,10^{-4}$, $h u_0 = -1$ and $\kappa = 12/23$ was used.

The problem was run to $x^0 = 360$. Two different time steps were employed $\Delta x^0 = 0.1875$ (identical to HSW) and $\Delta x^0 = 0.5$ (to demonstrate the method's performance with a larger Courant number). The results are shown in Fig. 15 and Table 7. The errors are at least an order of magnitude less than those of HSW. This should again mainly be attributed to the use of stationary extrapolation. The main test in this problem is finding the sonic point.

The two-dimensional flows that HSW show were not pursued here because they are especially geared towards accretion

**Table 6.** The maximal relative errors of the Roe solver and several HSW schemes for general relativistic dust accretion (test problem 5). HSW errors are as mentioned in their text. HSW refine the grid towards the horizon; the average spacing has been given here

| method | $\Delta t$ | $\Delta x^1$ | $\rho \frac{u^0}{\sqrt{-g^{00}}}$ | $\frac{u^1}{u^0}$ | $\rho e \frac{u^0}{\sqrt{-g^{00}}}$ | $p$ | $\rho h u_1 \frac{u^0}{\sqrt{-g^{00}}}$ |
|---|---|---|---|---|---|---|---|
| Roe | 0.5 | 1 | 0.06% | 0.01% | 0.68% | 0.69% | 0.04% |
| Roe | 0.1875 | 1 | 0.09% | 0.04% | 2.87% | 2.93% | 0.06% |
| HSW Wilson | 0.1875 | 0.67 | 3% | 1% | 30% | ? | ? |
| HSW Mono | 0.1875 | 0.67 | 1% | 2% | 10% | ? | ? |

**Table 7.** The maximal relative errors of the Roe solver and several HSW schemes for general relativistic Bondi-Hoyle accretion (test problem 6). HSW errors are as mentioned in their text. HSW refine the grid towards the horizon; the average spacing has been given here

| method | $\Delta t$ | $\Delta x^1$ | $\rho \frac{u^0}{\sqrt{-g^{00}}}$ | $\frac{u^1}{u^0}$ | $\rho e \frac{u^0}{\sqrt{-g^{00}}}$ | $p$ | $\rho h u_1 \frac{u^0}{\sqrt{-g^{00}}}$ |
|---|---|---|---|---|---|---|---|
| Roe | 0.5 | 1 | 0.15% | 0.17% | 0.26% | 0.27% | 0.04% |
| Roe | 0.1875 | 1 | 0.14% | 0.18% | 0.25% | 0.26% | 0.04% |
| HSW Wilson | 0.1875 | 0.67 | 10% | 2% | 23% | ? | ? |
| HSW Mono | 0.1875 | 0.67 | 3% | 2% | 16% | ? | ? |

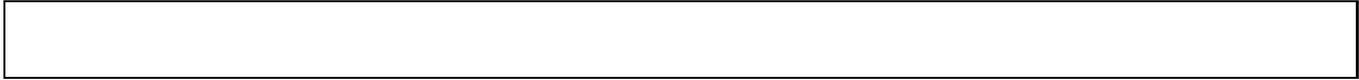

**Fig. 14.** Comparison of the numerical (crosses) and analytical solution for general relativistic dust accretion (test problem 5) on a 16 zone grid

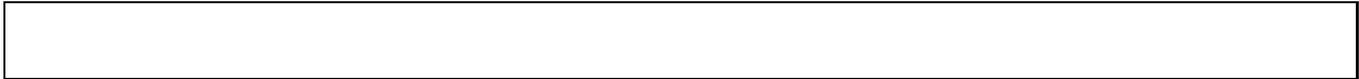

**Fig. 15.** Comparison of the numerical (crosses) and analytical solution for general relativistic Bondi-Hoyle accretion (test problem 6) on a 16 zone grid

flows, hardly allow quantitative comparisons and yet require substantial amounts of computational effort.

## 13. Discussion

In this paper we have presented a method for the numerical approximation of solutions of the equations of general relativistic fluid dynamics. The method is based on Roe's approximate Riemann solver for the classical Euler equations. A disadvantage of the method is that it is rather cumbersome and thus computationally expensive. However, it is physically appealing because of the direct physical meaning of characteristics as the paths along which information propagates. The economics is thus determined by whether the extra effort results in a sufficiently more accurate scheme. If so, the flow can sufficiently well be approximated on a coarser grid. This coarser grid pays back the extra work per cell, which becomes especially advantageous as calculations are done in more spatial dimensions. As shocks are quite difficult to reproduce numerically, it should be expected that the more violent flows benefit the most from the above scheme. Indeed, there is evidence from the above test problems that shows that the Roe scheme can be applied on a significantly coarser grid than the HSW schemes to obtain similarly accurate results. In addition the extremely relativistic shocks such as the ones that occur in the test problem of colliding special relativistic fluids (see Section 12), that until now required an implicit treatment, can be handled accurately by the explicit Roe solver.

The method presented in this paper has been successfully applied in two-dimensional flow simulations of interacting winds. It was discovered that a wide range of parameters pro-duces hydrodynamically collimated jets. The collimation mechanism was called "inertial confinement", because it is a mechanism by which jets form when a spherical hot wind blows outward through a thick accretion disk. Results for non relativistic planetary nebulae are given by Mellema et al. (1991), Icke et al. (1992), Mellema (1993), Mellema (1994), Mellema & Frank (1995); relativistic jets were calculated by Eulderink & Mellema (1994).

The code has not been yet applied in the context of active nuclei. But in cases of violent discontinuous flows we expect this to be worthwhile given the performance on the test problems.

As in Cartesian non relativistic hydrodynamics (Roe & Baines 1982; Roe 1986b), operator splitting to obtain quasi one-dimensional equations is an area of possible improvement for shock capturing schemes. In view of the above results it seems worthwhile to try to extend work on genuinely multi-dimensional characteristic-based methods, once proven to be successful in Cartesian non relativistic hydrodynamics, to the relativistic domain as well. The same holds for alternative implementations of the Roe solver which eliminate the possibility of unphysical fluxes.

## 14. Conclusions

1. The translation of Roe's concepts for an approximate Riemann solver from the non relativistic to the general relativistic domain is laborious but straightforward.
2. The implementation in a Minkowski metric completely analogous to the implementation in the Cartesian non relativistic hydrodynamics case presented by Roe (1986a), can be accomplished and is successful.
3. The implementation in GR or for curvilinear coordinates can be accomplished via a non-trivial subgrid model to deal with the source terms.
4. A source term splitting exists that allows algebraic integration of the curvature source terms.
5. The numerical scheme built around the GR Roe solver
   - yields excellent approximations to one-dimensional NR and SR shock tube problems,
   - is an explicit numerical scheme that overcomes the problems experienced with some other explicit schemes, when dealing with extremely relativistic shocks (examples up to $u^0 = 625$ have been given),
   - produces excellent approximations to the steady solution of GR radial accretion problems.

*Acknowledgements.* We wish to express our gratitude to Vincent Icke, Phil Roe, Tim de Zeeuw and Edwin Henneken for their useful comments on this article. We wish to thank Rob van Dijk who kindly provided a copy of his program that calculates analytical solutions to special relativistic Riemann problems, and Theo Verstrael who helped with the graphics. Finally, we owe an explanation to those who awaited the publication of this work since their interest was excited by the original 1988 preprint about the general relativistic Roe solver and its performance on wall-shocks: because of a career change, FE had to work on this article in his spare time.

## Appendix A: Determination of the primitive variables from the state

In this appendix three methods to determine the primitive variables given the conserved quantity vector are described: an analytical method, a one-dimensional Newton-Raphson method and a six-dimensional Newton-Kantorovich method.

### A..1. Analytical method

The first two ways of solving for the primitive variables amount to finding the root of a quartic polynomial. That polynomial results after substituting the mass and energy components of the state in terms of the primitive variables in the relation that is found if velocity normalization is used to replace the three momentum components by one relation:

$$\alpha_4 \xi^3(\xi - \eta) + \alpha_2 \xi^2 + \alpha_1 \xi + \alpha_0 = 0, \quad (A.1)$$

where

$$\xi \equiv \frac{\sqrt{-g_{\alpha\beta}T^{0\alpha}T^{0\beta}}}{\rho h u^0}, \quad (A.2)$$

$$\alpha_4 \equiv C_0^2 - 1,$$

$$\eta \equiv 2C\frac{\Gamma - 1}{\Gamma},$$

$$\alpha_2 \equiv -\frac{2-\Gamma}{\Gamma}\left(C_0^2 - 1\right) + 1 - C^2\left(\frac{\Gamma-1}{\Gamma}\right)^2,$$

$$\alpha_1 \equiv -2C\frac{\Gamma-1}{\Gamma^2},$$

$$\alpha_0 \equiv -\frac{1}{\Gamma^2}, \quad (A.3)$$

$$C_0 \equiv \frac{T^{00}}{\sqrt{g^{00}g_{\alpha\beta}T^{0\alpha}T^{0\beta}}},$$

$$C \equiv \frac{\rho u^0}{\sqrt{-g_{\alpha\beta}T^{0\alpha}T^{0\beta}}}. \quad (A.4)$$

Then $C_0 \geq 1$ because $u^0 u^0 \geq -g^{00}$, and $0 < C < 1$ because pressure is positive. Therefore the coefficient of the fourth power is non-negative. It may also be deduced that physically allowed values for $\xi$ are in the range

$$0 < \sqrt{\frac{2-\Gamma}{\Gamma}} < \xi < 1. \quad (A.5)$$

Proof that precisely two real roots for all allowed coefficients exist, and that only one of these roots obeys the physical constraint that it is positive, is presented first.

The roots of the polynomial Eq. (A.1) can also be expressed as the intersection points of two polynomials:

$$(C_0^2 - 1)\xi^2 \left(\xi - \left[\frac{\Gamma-1}{\Gamma}C - \left(\frac{2-\Gamma}{\Gamma} + \left\{\frac{\Gamma-1}{\Gamma}\right\}^2 C^2\right)^{1/2}\right]\right)$$
$$\left(\xi - \left[\frac{\Gamma-1}{\Gamma}C + \left(\frac{2-\Gamma}{\Gamma} + \left\{\frac{\Gamma-1}{\Gamma}\right\}^2 C^2\right)^{1/2}\right]\right)$$
$$= -\left(\left[1 + \frac{\Gamma-1}{\Gamma}C\right]\xi + \frac{1}{\Gamma}\right)\left(\left[1 - \frac{\Gamma-1}{\Gamma}C\right]\xi - \frac{1}{\Gamma}\right). \quad (A.6)$$

The right-hand side polynomial is negative everywhere except on the interval between its roots. Thus if $C_0 = 1$ the existence of precisely one intersection in the physically acceptable region is easily proven. If $C_0 > 1$, the polynomial on the left-hand side has one root that is smaller than zero, a double root at $\xi = 0$ and one root $\xi > 0$. This polynomial is positive everywhere except for the region between its two roots. From Eq. (A.6) it may now easily be deduced that there is one root in the physically acceptable region

$$\left(\frac{\Gamma-1}{\Gamma}C + \left\{\frac{2-\Gamma}{\Gamma} + \left(\frac{\Gamma-1}{\Gamma}\right)^2 C^2\right\}^{\frac{1}{2}}, \frac{1}{\Gamma-(\Gamma-1)C}\right] \quad (A.7)$$

and one other root in the interval

$$\left[\frac{-1}{\Gamma+(\Gamma-1)C}, \frac{\Gamma-1}{\Gamma}C - \left\{\frac{2-\Gamma}{\Gamma} + \left(\frac{\Gamma-1}{\Gamma}\right)^2 C^2\right\}^{\frac{1}{2}}\right). \quad (A.8)$$

Thus precisely one physical solution exists of the quartic Eq. (A.1).

In the classical limit $C \uparrow 1$ and $C_0 \downarrow 1$, which causes $\alpha_4$ to vanish. The two roots that remain are $\xi \to 1$ and $\xi \to -1/(2\Gamma-1)$. The physical constraints imply that the first root is the ratio sought, because the latter is smaller than zero. The root formula presented below does indeed pick this correct solution in the classical limit. Because the root of the polynomial is given below as a continuous function of its coefficients and the coefficients in their turn are continuous functions of the conserved variables, this means that the proper root is selected in the relativistic case also since such a root was proven to exist.

Having selected the one root that satisfies the physical constraints it can be determined by the exact algebraic quartic root formula. However, special care must be taken, because the coefficients of the fourth and third powers of the polynomial tend to zero in the NR limit. The familiar expressions for the root of a quartic equation should thus be rewritten in a form that can accommodate this limit behaviour.

The selected root $\xi$ is found to be

$$\xi = \tfrac{1}{2}\left(\gamma_1 + \sqrt{\gamma_1^2 - 2\gamma_0}\right), \quad (A.9)$$

where

$$\gamma_0 = 4\alpha_0 \left((\alpha_2 - \alpha_4\delta) + [(\alpha_2 - \alpha_4\delta)^2 - 4\alpha_0\alpha_4]^{\frac{1}{2}}\right)^{-1},$$
$$\gamma_1 = \tfrac{1}{2}\left(\eta + \sqrt{\eta^2 - 4\delta}\right),$$
$$\delta = \frac{1}{\nu_1\nu_2}\left[\beta_3(\nu_1 + \nu_2) + \frac{2\beta_4^2(\sigma + \tau + \tfrac{1}{3}\alpha_2)}{\sigma^2 + \sigma\tau + \tau^2}\right]. \quad (A.10)$$

The auxiliary symbols used are

$$\nu_1 = \sigma^2 + \tfrac{1}{3}\sigma\alpha_2 + \tfrac{1}{9}\alpha_2^2,$$
$$\nu_2 = \tau^2 + \tfrac{1}{3}\tau\alpha_2 + \tfrac{1}{9}\alpha_2^2,$$
$$\sigma = \left(\tfrac{1}{27}\alpha_2^3 - \alpha_4\beta_3 + \sqrt{\alpha_4}\beta_4\right)^{\frac{1}{3}},$$
$$\tau = \left(\tfrac{1}{27}\alpha_2^3 - \alpha_4\beta_3 - \sqrt{\alpha_4}\beta_4\right)^{\frac{1}{3}},$$
$$\beta_3 = \tfrac{1}{2}\left(\tfrac{1}{3}\alpha_2\beta_1 + \beta_0\right),$$
$$\beta_4 = \left[\tfrac{1}{27}\left(-\beta_0\alpha_2^3 + \alpha_4^2\beta_1^3 - \tfrac{1}{4}\alpha_4\beta_1^2\alpha_2^2\right) + \alpha_4\beta_0\left(\tfrac{1}{4}\beta_0 + \tfrac{1}{6}\beta_1\alpha_2\right)\right]^{\frac{1}{2}},$$
$$\beta_0 = 4\alpha_0\alpha_2 - \alpha_1^2 - \eta^2\alpha_0\alpha_4,$$
$$\beta_1 = -(\eta\alpha_1 + 4\alpha_0). \quad (A.11)$$

Once $\xi$ is known, one finds by combining the expressions for $C$, $C_0$, $T^{00}$ and using $hC\xi = 1$ that the time component of the four-velocity is

$$u^0 = \tfrac{1}{2}\sqrt{-g^{00}}\left\{C_0\xi + \left[C_0^2\xi^2 + 4\frac{\Gamma-1}{\Gamma}(1 - C\xi)\right]^{\frac{1}{2}}\right\}. \quad (A.12)$$

The density is now easy to determine:

$$\sqrt{-g}\rho = \frac{(\sqrt{-g}\rho u^0)}{u^0}. \quad (A.13)$$

The enthalpy, pressure and velocities are then determined by

$$\sqrt{-g}\rho h = \frac{(\sqrt{-g}T^{00})}{\sqrt{-g^{00}}C_0\xi u^0},$$
$$p = \frac{\Gamma-1}{\Gamma}\rho h(1 - C\xi),$$
$$\frac{u^j}{u^0} = \frac{(\sqrt{-g}T^{0j}) - \sqrt{-g}pg^{0j}}{\sqrt{-g}\rho h u^0 u^0}. \quad (A.14)$$

### A..2. One-dimensional Newton-Raphson method

The above procedure is computationally rather expensive because of the many extra variables and exponentiations involved, and because special care that must be taken when numerically evaluating the cubic roots. An alternative way to find the ratio is based on an iterative procedure. The obvious choice is to use Newton's method on the polynomial Eq. (A.4).

With some effort it may be proven that the derivative of the polynomial of Eq. (A.4) is always at least $2/\Gamma$ in its physical root, thus guaranteeing quadratic convergence of the Newton method. This is the main reason to choose $\xi$ and not some other variable. With the starting value that is asymptotically correct in the NR limit (i.e. as $C, C_0 \to 1$)

$$\xi_{start} = 1, \qquad (A.15)$$

only five iterations of the Newton method where required to obtain a relative accuracy of $10^{-16}$ in all test problems. This is a lot less expensive than the algebraic formula presented above.

### A..3. Six-dimensional Newton-Kantorovich method

The third alternative to solve for the primitive variables is to use a Newton-Kantorovich method on a system of equations. This method is presented below in a form that yields

$$K \equiv (\sqrt{-g}\rho h)^{\frac{1}{2}}, \qquad (A.16)$$

and the parameter vector, defined as

$$w \equiv (w^\alpha, w^4) \equiv (w^0, w^1, w^2, w^3, w^4), \qquad (A.17)$$

where
$$w^\alpha \equiv K u^\alpha,$$
$$w^4 \equiv K \frac{p}{\rho h}. \qquad (A.18)$$

Simple explicit expressions exist for the primitive variables in terms of $K$ and the elements of the parameter vector. To determine $K$ and the parameter vector given the state vector of a relativistic fluid one must solve a coupled set of equations. The first equation gives a relation between the variables, and the other five determine the relation of the parameter vector to the state:

$$\begin{pmatrix} 0 \\ 0 \\ 0 \\ 0 \\ 0 \\ 0 \end{pmatrix} = \begin{pmatrix} \tilde{\Delta} \\ \Delta \\ \Delta^0 \\ \Delta^1 \\ \Delta^2 \\ \Delta^3 \end{pmatrix} \equiv \begin{pmatrix} -\frac{1}{2}(K^2 + w^\alpha w_\alpha) \\ \sqrt{-g}\rho u^0 - w^0 \left( K - \frac{\Gamma}{\Gamma-1} w^4 \right) \\ \sqrt{-g} T^{00} - (w^0 w^0 + K w^4 g^{00}) \\ \sqrt{-g} T^{01} - (w^0 w^1 + K w^4 g^{01}) \\ \sqrt{-g} T^{02} - (w^0 w^2 + K w^4 g^{02}) \\ \sqrt{-g} T^{03} - (w^0 w^3 + K w^4 g^{03}) \end{pmatrix}. \quad (A.19)$$

Given a starting value for the parameter vector, solve for the increases in the elements of the parameter vector that would make the right-hand side equal to 0 if the equations where linear in the chosen variables. These increases are given by

$$\begin{pmatrix} K & w_0 & w_1 & w_2 & w_3 & 0 \\ w^0 & K - \frac{\Gamma}{\Gamma-1} w^4 & 0 & 0 & 0 & -\frac{\Gamma}{\Gamma-1} w^0 \\ w^4 g^{00} & 2 w^0 & 0 & 0 & 0 & K g^{00} \\ w^4 g^{01} & w^1 & w^0 & 0 & 0 & K g^{01} \\ w^4 g^{02} & w^2 & 0 & w^0 & 0 & K g^{02} \\ w^4 g^{03} & w^3 & 0 & 0 & w^0 & K g^{03} \end{pmatrix} \begin{pmatrix} \Delta K \\ \Delta w^0 \\ \Delta w^1 \\ \Delta w^2 \\ \Delta w^3 \\ \Delta w^4 \end{pmatrix}$$

$$= \begin{pmatrix} \tilde{\Delta} \\ \Delta \\ \Delta^0 \\ \Delta^1 \\ \Delta^2 \\ \Delta^3 \end{pmatrix}. \qquad (A.20)$$

This linear set of equations may be solved to give

$$\begin{pmatrix} \Delta K \\ \Delta w^0 \end{pmatrix} = \frac{1}{\det} \begin{pmatrix} K^2 - \frac{\Gamma}{\Gamma-1} \left( K w^4 - 2 \frac{w^0 w^0}{g^{00}} \right) \\ -w^0 \left( K + \frac{\Gamma}{\Gamma-1} w^4 \right) \end{pmatrix}$$
$$\begin{pmatrix} \frac{w^\alpha w_\alpha}{w^0} - 2 \frac{w^0}{g^{00}} \\ K \end{pmatrix} \begin{pmatrix} \tilde{\Delta} - \frac{w_i}{w^0} \left( \Delta^i - \frac{g^{0i}}{g^{00}} \Delta^0 \right) \\ K \Delta + \frac{\Gamma}{\Gamma-1} \frac{w^0}{g^{00}} \Delta^0 \end{pmatrix}, \quad (A.21)$$

where the determinant is

$$\det = K \left( K^2 - \frac{\Gamma}{\Gamma-1} \left[ K w^4 - 2 \frac{w^0 w^0}{g^{00}} \right] \right)$$
$$+ \left( K + \frac{\Gamma}{\Gamma-1} w^4 \right) \left( w^\alpha w_\alpha - 2 \frac{w^0 w^0}{g^{00}} \right). \qquad (A.22)$$

Given these two increases the others are

$$\Delta w^1 = \frac{1}{w^0} \left( \Delta^1 - \frac{g^{01}}{g^{00}} \Delta^0 \right) - \left( w^1 - 2 w^0 \frac{g^{01}}{g^{00}} \right) \frac{\Delta w^0}{w^0},$$

$$\Delta w^2 = \frac{1}{w^0} \left( \Delta^2 - \frac{g^{02}}{g^{00}} \Delta^0 \right) - \left( w^2 - 2 w^0 \frac{g^{02}}{g^{00}} \right) \frac{\Delta w^0}{w^0},$$

$$\Delta w^3 = \frac{1}{w^0} \left( \Delta^3 - \frac{g^{03}}{g^{00}} \Delta^0 \right) - \left( w^3 - 2 w^0 \frac{g^{03}}{g^{00}} \right) \frac{\Delta w^0}{w^0},$$

$$\Delta w^4 = \frac{1}{K g^{00}} \Delta^0 - w^4 \frac{\Delta K}{K} - 2 \frac{w^0}{g^{00}} \frac{\Delta w^0}{K}. \qquad (A.23)$$

These increases determine a new parameter vector. Because Eq. (A.19) is in fact non-linear these steps must be iterated to obtain convergence.

In principle one can substitute the first relation of Eq. (A.19) in the other five to eliminate $K$. The reason this is not done and $K$ is treated as an independent variable instead is that one circumvents the evaluation of the square root. This is not only faster but also avoids problems that otherwise arise if convergence goes through a region of negative $K^2$.

## Appendix B: Example of a super-resolution problem

In this appendix we will give a specific example of how a false assumption about the subgrid model can influence the flow. Consider NR stationary radial outflow with supersonic velocity $u^1$. The NR limit of stationary extrapolation in the 1-direction yields that the mass flux

$$D = \sqrt{-g}\rho u^1, \qquad (B.1)$$

the Bernoulli constant

$$E \equiv \frac{1}{2} g_{ij} u^i u^j + \frac{\Gamma}{\Gamma-1} \frac{p}{\rho}, \qquad (B.2)$$

the transverse velocity

$$u_k, \qquad k = 2, 3 \tag{B.3}$$

and the entropy

$$\kappa = \frac{p}{\rho^\Gamma} \tag{B.4}$$

are invariant. In the NR limit the density $\rho$ is to be solved from

$$E = \tfrac{1}{2} \left[ \frac{1}{g^{11}} \left( \frac{D}{\sqrt{-g}\rho} \right)^2 - \frac{1}{g^{11}} \left( g^{1k} u_k \right)^2 + g^{kl} u_k u_l \right]$$
$$+ \frac{\Gamma}{\Gamma - 1} \kappa \rho^{\Gamma - 1}, \qquad k, l = 2, 3. \tag{B.5}$$

In spherical coordinates (radius $r$, polar angle $\theta$ and azimuthal angle $\phi$), the density at radius $r$ is thus determined by

$$E = \tfrac{1}{2} \left( \frac{D}{r^2 \sin\theta \rho} \right)^2 + \frac{\Gamma}{\Gamma - 1} \kappa \rho^{\Gamma - 1}. \tag{B.6}$$

Hence

$$\frac{d \ln \rho}{d \ln r} = 2 \frac{M^2}{1 - M^2}, \tag{B.7}$$

where

$$M^2 = \frac{\rho u^1 u^1}{\Gamma p}. \tag{B.8}$$

Now consider the same flow at the same point in space in cylindrical coordinates (indicated by $\tilde{\ }$): $\tilde{r} = r \sin\theta$, $\tilde{z} = r \cos\theta$ and $\tilde{\phi} = \phi$. The total velocity is split in a polar component $\tilde{u}^1 = u^1 \sin\theta$ and a component parallel to the axis of symmetry $\tilde{u}^2 = u^1 \cos\theta$. Extrapolation to a different $r$ is done in two steps. A step in the 2-direction that leaves all quantities the same and a step in the 1-direction that yields

$$\tilde{E} = \tfrac{1}{2} \left( \frac{\tilde{D}}{\tilde{r}\rho} \right)^2 + \tfrac{1}{2} \left( \tilde{u}^2 \right)^2 + \frac{\Gamma}{\Gamma - 1} \tilde{\kappa} \rho^{\Gamma - 1}, \tag{B.9}$$

where $\tilde{D} = \frac{D}{r} \sin\theta$, $\tilde{\kappa} = \kappa$ and $\tilde{E} = E$. Hence

$$\frac{d \ln \rho}{d \ln \tilde{r}} = \frac{\tilde{M}^2}{1 - \tilde{M}^2}, \tag{B.10}$$

where

$$\tilde{M}^2 \equiv \frac{\rho \tilde{u}^1 \tilde{u}^1}{\Gamma p} = M^2 \sin^2 \theta. \tag{B.11}$$

Apart from the different pre-factor there is an even more serious problem with these formulae: there is a pole for sonic flow. In the spherical coordinates where the velocity is supersonic the density decreases away from the centre. In cylindrical coordinates however, the subgrid model has a supersonic *or subsonic* 1-velocity depending on the projection of the total velocity, which in its turn depends on the position on the grid via $\theta$. In case it is subsonic the density *increases* outward from the axis. This is not just a sign error (which is bad enough). Because of the pole the difference is extreme at the position on the grid where the *projected* velocity is near sonic, i.e. $\sin\theta = 1/M$. However, as the flux varies smoothly through $M = 1$ it may not be affected quite so badly. Nevertheless, in principle instabilities can result from the false assumption of the subgrid model that stationary spherical flow can be approximated by stationary cylindrical flow. In numerical experiments on multi-dimensional flows one should be aware of these possible instabilities.

**Appendix C: Implementations for general relativistic flow**

In this appendix first the method that results if operator splitting as discussed in Sect. 3 is combined with an arbitrary extrapolation of Sect. 6, is described. Next, it is shown how to arrive at the method presented in Sect. 8. Functions that are known a-priori are collectively called "known functions". The "unknowns" are functions of a minimal set that needs to be known in addition to the known functions to completely describe the state and that need to be determined by integration of the partial differential equations. In the GRH setting one may think of the primitive variables.

*C..1. Basic method*

Straight application of operator splitting as discussed in Sect. 3 and any source term splitting as discussed in Sections 3 and 6 results in the following implementation:

1. Given the state $[F^0]_i^n$ determine the unknowns $u_i^n$.
2. Extrapolate the unknowns to the intermediate time level $[x^0]_{n+\frac{1}{2}-}$ assuming

$$F_{,0}^0 = T. \tag{C.1}$$

Call the unknowns $u_i^{n+\frac{1}{2}-}$. For this extrapolation the unknowns $u$ change with time as

$$u_{,0} = (F_u^0)^{-1}(T - F_g^0 g_{,0}). \tag{C.2}$$

3. Extrapolate the unknowns at cell centre $[x^1]_i$ to the interfaces at $[x^1]_{i-\frac{1}{2}}$ and $[x^1]_{i+\frac{1}{2}}$, assuming

$$F_{,1}^1 = X. \tag{C.3}$$

The resulting unknowns are called $u_{i-\frac{1}{2}+}^{n+\frac{1}{2}-}$ and $u_{i+\frac{1}{2}-}^{n+\frac{1}{2}-}$ respectively. For this extrapolation the unknowns $u$ change with position as

$$u_{,x} = (F_u^1)^{-1}(X - F_g^1 g_{,1}). \tag{C.4}$$

4. At each interface (e.g. at $[x^1]_{i-\frac{1}{2}}$) solve the (approximate) Riemann problem defined by the local known functions and the unknowns immediately left and right of that interface. This yields the unknowns $[u]_{i-\frac{1}{2}-}^{n+\frac{1}{2}+}$ and $[u]_{i-\frac{1}{2}+}^{n+\frac{1}{2}+}$.

5. Using Eq. (C.4), extrapolate the unknowns $[u]_{i-\frac{1}{2}+}^{n+\frac{1}{2}+}$ and $[u]_{i+\frac{1}{2}-}^{n+\frac{1}{2}+}$ from the interface to the cell centre to form $[u]_{i-}^{n+\frac{1}{2}+}$ and $[u]_{i+}^{n+\frac{1}{2}+}$, respectively.
6. Determine the fluxes that belong to these unknowns: $[F^1]_{i-}^{n+\frac{1}{2}+}$ and $[F^1]_{i+}^{n+\frac{1}{2}+}$.
7. Update the state according to

$$[F^0]_i^{n+\frac{1}{2}+} = [F^0]_i^{n+\frac{1}{2}-} - \frac{\Delta x^0}{\Delta x^1}\left([F^1]_{i+}^{n+\frac{1}{2}+} - [F^1]_{i-}^{n+\frac{1}{2}+}\right). \quad (C.5)$$

8. Given the state $[F^0]_i^{n+\frac{1}{2}+}$ determine the unknowns $u_i^{n+\frac{1}{2}+}$.
9. Using Eq. (C.2), extrapolate these unknowns in time, to form the final update $[F^0]_i^{n+1}$.

Since the update sequence begins and ends with an integration over time, these integrations may be combined to yield $[F^0]_i^{n+\frac{1}{2}-}$ immediately from $[F^0]_i^{n-\frac{1}{2}+}$. Thus one can dispense with steps (1) and (9) if one is only after a next update.

In case there is no source term and no dependence on known functions the old formulae are automatically recovered, because the extrapolations become trivial and only steps (4), (7), and (8) are then necessary. As is shown in Appendix D, the extrapolation in time must be at least *second* order accurate, while the extrapolation in space only has to be *first* order accurate. This difference in required accuracy arises because in the spatial direction information can be obtained from neighbours on both sides whereas in the time direction one has only the past to predict the future. More specifically, the second order spatial extrapolation terms cancel in the flux difference. In general extrapolation serves to determine the unknowns at a new location in spacetime. Extrapolations that yield combinations of the known and unknown functions must therefore be followed by a step in which the unknowns are determined. In any case, a procedure as described in Appendix A is necessary to determine the unknowns after the changes to the state of step (7).

The total amount of work per cell and per time step in one spatial direction is:

1 time extrapolation
4 space extrapolations
5 unknowns determinations after extrapolation
1 (approximate) Riemann problem
2 unknowns determinations after the Riemann solver
1 unknowns determination from the state.

For smooth flow this implementation is second order accurate both in space and in time. This is obvious if one realizes that the recipe above consists of a second order accurate integration in time and a second order accurate integration in space combined in such a way as to yield a second order result. Alternatively one may prove that the methods are second order accurate by writing out the nested dependencies. This is done in Appendix D.

### C..2. Algebraic extrapolation

A special case of the above scheme results if the algebraic subgrid model of Sect. 6 is chosen. After stationary/homogeneous extrapolation the local density is available and consequently determining the new unknowns is trivial. The use of analytical solutions for extrapolation makes the GRH scheme as robust as the NRC scheme. It allows the use of the same methods that guarantee a physical prediction in CH to guarantee an physical prediction in the GRH case. More specifically, if one uses algebraic extrapolation, a flux limiter that produces physical interface states for the Riemann solver and a Riemann solver that yields a physical interface state (such as the exact solver), a physical approximation for the updated state is guaranteed even in GRH! This property is due to two facts: first, the extrapolation and the Riemann solver predict physically allowed states. Second, only vectors that exist at the same place and time are combined to yield an update. Notice that if the extrapolation equations are integrated numerically, numerical inaccuracies can still cause the result to be unphysical.

Advantages of this implementation are: one-dimensional stationary solutions are automatically recognized as such, specific use is made of the coordinate symmetries, and the concept of operator splitting in the coordinate directions is strictly applied. Hence, changes in the state due to changes in the source term after the passage of a discontinuity are consistently taken into account. The operator splitting results in equations to which physical solutions exist.

The total work involved is

1 time extrapolation
4 space extrapolations
0 unknowns determinations after extrapolation
1 (approximate) Riemann problem
2 unknowns determinations after the Riemann solver
1 unknowns determination from the state.

### C..3. Determination of the unknown variables from the flux

The use of extrapolation that requires knowledge of the unknowns such as in the previous subsection in combination with a Roe solver has a disadvantage: a Roe solver only yields an interface flux and not the interface state or the interface unknowns (see Sect. 4). Stationary extrapolation of the flux without knowledge of the primitive variables fails because only the constants $D$ and $Dhu_\epsilon$ can easily be determined from a given flux, but not the entropy. Yet, this entropy is required to find the cell centre flux.

The unknowns can be determined uniquely from the state, but in general not from the flux. If

$$D^2 + g_{\alpha\beta}(\sqrt{-g}T^{1\alpha})(\sqrt{-g}T^{1\beta}) < 0, \quad (C.6)$$

there exist two different entropies that are consistent with a given flux. This is related to the possibility of a stationary shock: the entropy may change while the flux remains constant. Thus supersonic and subsonic primitive variables may exist that yield the same flux. Different branches may have be chosen on either side of an interface, because of the possibility of a steady shock. If the mass flux $D = 0$ the flow is even less determined: $u^1 = 0$ but the three transverse constants $hu_\epsilon$ and the entropy are only related through the pressure at the place where the flux is known.

There are two ways to resolve this problem. One can either use additional information to choose a branch or one can try to circumvent having to evaluate the unknowns from the flux. In this subsection the former path is taken and in the remaining subsections of this appendix the latter path is explored.

In view of the non-uniqueness one must use extra information, such as the initial states that entered the Riemann problem, to determine unknowns from fluxes. The unknowns can be different on either side of the interface due to the presence of a stationary contact discontinuity or shock. Therefore a physical solution to the mathematical problem above is to use the entropy and transverse constants determined by the last available state on either side of the cell interface.

A possible way to determine an interface state instead of an interface flux by a (first order) Roe type method is

$$[F^0]_{i+\frac{1}{2}-}^{n+\frac{1}{2}+} = [F^0]_{i+\frac{1}{2}-}^{n+\frac{1}{2}-} + \sum\nolimits^{(-)} a_k e_k,$$
$$[F^0]_{i+\frac{1}{2}+}^{n+\frac{1}{2}+} = [F^0]_{i+\frac{1}{2}+}^{n+\frac{1}{2}-} - \sum\nolimits^{(+)} a_k e_k, \qquad (C.7)$$

where $\sum^{(-)}$ and $\sum^{(+)}$ denote summation over all indices of negative and positive characteristic speeds, respectively.

Different interface states are predicted immediately left and right of the interface, because there *exists* no single interface state if the states immediately left and right are separated by a shock or contact discontinuity that has no velocity with respect to the grid. Rather there are *two* distinct states immediately left and right of an interface in such a situation. These states cannot be used directly to predict the interface flux, because the two approximate interface states do not necessarily correspond to the same flux. Hence, such a scheme would not be conservative. Yet, they can be used to find on each side the branch of the primitive variables that correspond to the Roe flux. Notice that waves with characteristic speed zero do not affect the interface states. Picking the wrong entropy branch near sonic points does not cause large errors in the flux.

### C..4. Time extrapolation only

Several alternatives exist to prevent extrapolation of the interface flux. To minimize the amount of work one may be tempted to choose $T = S$ and $X = 0$. This corresponds to a subgrid model that prescribes a constant flux. Spatial extrapolation is thus trivial. After the extrapolation of cell interface fluxes to the cell centre no unknowns need to be determined because the fluxes suffice. Hence the total amount of work is

  1 time extrapolation
  0 space extrapolations
  3 unknowns determinations after extrapolation
  1 (approximate) Riemann problem
  0 unknowns determinations after the Riemann solver
  1 unknowns determination from the state.

However, there are also drawbacks. One-dimensional stationary solutions are not automatically recognized as such. Unphysical interface quantities are likely to occur if the metric is space dependent. No specific use of the coordinate symmetries is made. After the addition of the flux difference an interface state must be evaluated while curvature terms have not yet been evaluated, which again could lead to unphysical intermediate results, especially if discontinuities cross the cell.

In this case the unknowns that enter the Riemann problem must be determined after the extrapolation of the flux from the cell centre to the cell interface. In this case the uniqueness issues may be solved by selecting a branch after numerically integrating the extrapolation equation for the unknowns.

### C..5. Centre-centre extrapolation

Another way to avoid the determination of the primitive variables from the flux for solvers that only yield a flux is to combine the extrapolation before and after the approximate Riemann problem to just an extrapolation before the approximate Riemann problem: the unknowns at the centre of cell $i - 1$ are extrapolated to the centre of cell $i$. There these unknowns and the unknowns that correspond to the centre of cell $i$ are combined by an (approximate) Riemann solver to yield a flux. Likewise a flux can be formed by the Riemann solver based on the extrapolation of the unknowns of cell centre $i + 1$ and the unknowns at $[x^1]_i$. The two fluxes determine the update of the state in cell $i$. Because the above extrapolation is not symmetric, two Riemann problems must be solved per interface instead of one.

At first one might be tempted to consider the approximate Riemann problems as Riemann problems at $[x^1]_i$, because that is where the unknowns exist. However to make the scheme second order accurate the transport speeds that enter the recipe via the Courant number must be evaluated at the interface as must the eigenvectors. Also, a scheme that takes these quantities at the cell centre is not conservative for equations without a source term: the eigenvalues and/or the eigenvectors of systems of equations depend on the state (and possibly the metric). Thus the projection of the same flux on the local eigenvectors is in general different from one place to another. Consequently, a part of the flux that is considered upwind at $[x^1]_i$ may not be considered downwind at $[x^1]_{i-1}$. Therefore the flux that enters cell $i$ may not be the same as the flux leaving cell $i-1$. To resolve this problem one should project the above flux difference on the eigenvectors at the interface, although the fluxes are defined at the centre of cell $i$ and have no physical meaning on the interface. To obtain the unknowns at the interface the cell centre unknowns must be extrapolated to the interface as well as to the cell centre of their neighbour's cell centre. Using the *interface* characteristics guarantees that if there is no source term and consequently the flux difference vector at $[x^1]_{i-1}$ is the same as the flux difference vector at $[x^1]_i$, the approximations to their state differences come out the same too. Thus the calculated interface flux is the same in those cases. This guarantees a conservative scheme if no source terms are present.

However, if there is a source term in one or more but not in all equations of a set, the two flux differences above are no longer the same (as a vector), and it is possible that the conserved quantities corresponding to the equations without source term are not conserved by this scheme. This may not be as bad as it sounds, since the proper treatment of shocks is the strongest motive for placing emphasis on conservation and shocks are likely to be treated well by the method; nevertheless, a conservative scheme is preferable

### C..6. Approximate extrapolation

To save the extra work involved in the algebraic extrapolations or to take external source terms into account, one may consider numerical integration of the source term. Numerical integration of the spatial extrapolation equation is straightforward as it only needs to be integrated to first order accuracy. The time extrapolation has to be second order. The scheme below is especially apt to deal with numerical extrapolation:

1. Given the state $[F^0]_i^{n-\frac{1}{2}+}$ determine the unknowns $u_i^{n-\frac{1}{2}+}$.
2. Calculate the source term $X_i^{n-\frac{1}{2}+}$.
3. Extrapolate the unknowns to the time level $[x^0]_n$

$$[u]_i^n = [u]_i^{n-\frac{1}{2}+} + \tfrac{1}{2}\Delta x^0 \left[(F_u^0)^{-1}(T - F_g^0 g_{,0})\right]_i^{n-\frac{1}{2}+}. \quad (C.8)$$

4. Extrapolate the unknowns to the intermediate time level $[x^0]_{n+\frac{1}{2}}$ via

$$[u]_i^{n+\frac{1}{2}-} = [u]_i^{n-\frac{1}{2}+} + \Delta x^0 \left[(F_u^0)^{-1}(T - F_g^0 g_{,0})\right]_i^n. \quad (C.9)$$

5. Calculate the corresponding state $[F^0]_i^{n+\frac{1}{2}-}$ and spatial source term $X_i^{n+\frac{1}{2}-}$.
6. Extrapolate the unknowns from the cell centre at $[x^1]_i$ to the interfaces at $[x^1]_{i-\frac{1}{2}}$ and $[x^1]_{i+\frac{1}{2}}$, assuming

$$[u]_{i-\frac{1}{2}+}^{n+\frac{1}{2}-} = [u]_i^{n+\frac{1}{2}-} - \tfrac{1}{2}\Delta x^1 \left[(F_u^1)^{-1}(X - F_g^1 g_{,1})\right]_i^{n+\frac{1}{2}-},$$

$$[u]_{i+\frac{1}{2}-}^{n+\frac{1}{2}-} = [u]_i^{n+\frac{1}{2}-} + \tfrac{1}{2}\Delta x^1 \left[(F_u^1)^{-1}(X - F_g^1 g_{,1})\right]_i^{n+\frac{1}{2}-}.$$
$$(C.10)$$

7. Calculate the states $[F^0]_{i-\frac{1}{2}+}^{n+\frac{1}{2}-}$ and $[F^0]_{i+\frac{1}{2}-}^{n+\frac{1}{2}-}$ immediately right and left of the interfaces of each cell.
8. At each interface (e.g. at $[x^1]_{i-\frac{1}{2}}$) solve the (approximate) Riemann problem defined by the local known functions and the unknowns immediately left and right of that interface. This yields the flux $[F^1]_{i-\frac{1}{2}}^{n+\frac{1}{2}+}$.
9. Extrapolate the fluxes to the cell centre, based on the *old* source term:

$$[F^1]_{i-}^{n+\frac{1}{2}+} = [F^1]_{i-\frac{1}{2}}^{n+\frac{1}{2}} + \tfrac{1}{2}\Delta x^1 [X]_i^{n+\frac{1}{2}-},$$

$$[F^1]_{i+}^{n+\frac{1}{2}+} = [F^1]_{i+\frac{1}{2}}^{n+\frac{1}{2}} - \tfrac{1}{2}\Delta x^1 [X]_i^{n+\frac{1}{2}-}. \quad (C.11)$$

10. Update the state via

$$[F^0]_i^{n+\frac{1}{2}+} = [F^0]_i^{n+\frac{1}{2}-} - \frac{\Delta x^0}{\Delta x^1}\left([F^1]_{i+}^{n+\frac{1}{2}+} - [F^1]_{i-}^{n+\frac{1}{2}+}\right)$$
$$+ \Delta x^0 \left([X]_i^{n-\frac{1}{2}+} - [X]_i^{n-\frac{1}{2}-}\right). \quad (C.12)$$

For this method the extrapolations are so cheap that they are ignored. The amount of work is:

0 time extrapolations
0 space extrapolations
3 unknowns determinations after extrapolation
1 (approximate) Riemann problem
0 unknowns determinations after the Riemann solver
1 unknowns determination from the state.

This method involves less work than the alternatives mentioned above, but it has the disadvantage that one-dimensional stationary solutions are no longer recognized and that the linear extrapolation in space and time may not yield physically allowed states. Particularly a discontinuity may cause the intermediate result to be unphysical.

In the update only the spatial flux differences enter; the fluxes themselves do not. Therefore for a *second order* scheme the estimates for the fluxes immediately left and right of an interface are sufficiently well approximated by *first order* extrapolation from the cell centre, since the second order terms cancel in the flux difference. This allows one to use first order extrapolation. An implementation based on approximate extrapolation was used to obtain the results of Mellema et al. (1991), Icke et al. (1992), Mellema (1993), Eulderink & Mellema (1994), Mellema (1994), and Mellema & Frank (1995).

### C..7. Mixed extrapolation

The method that was used in this paper is based on the source term splitting of Sect. 6 and on algebraic extrapolation from the cell centre to the interface and linear extrapolation followed by a source term correction in the opposite direction. This may seem unbalanced but it offers the advantage that a one-dimensional stationary solution is recognized, while at the same time the difficult step of determining unknowns from the Roe flux is evaded. The amount of work is then

1 time extrapolations
2 space extrapolations
0 unknowns determinations after extrapolation
1 (approximate) Riemann problem
0 unknowns determinations after the Riemann solver
1 unknowns determination from the state.

Time extrapolation is trivial in all examples mentioned in this paper, because none of them have a time dependent metric.

### Appendix D: Proof of second order accuracy

In this appendix it is proven that for smooth flow the method presented in Appendix C is second order accurate both in space and time, by writing out all the nested dependencies. Only terms that are relevant to prove that the method is second order accurate are kept.

The update for the state is

$$[F^0]_i^{n+\frac{1}{2}+} \equiv [F^0]_i^{n+\frac{1}{2}-} - \frac{\Delta x^0}{\Delta x^1}\left([F^1]_{i+}^{n+\frac{1}{2}+} - [F^1]_{i-}^{n+\frac{1}{2}+}\right),$$

$$[F^0]_i^{n+\frac{1}{2}-} \approx [F^0]_i^n + \frac{1}{2}\Delta x^0 [T]_i^n$$
$$+ \frac{1}{8}(\Delta x^0)^2 \left[T_g g_{,0} + T_u(F_u^0)^{-1}\left(T - F_g^0 g_{,0}\right)\right],$$

$$[F^1]_{i-}^{n+\frac{1}{2}+} \approx [F^1]_{i-\frac{1}{2}}^{n+\frac{1}{2}+} + \frac{1}{2}\Delta x^1 [X]_{i-\frac{1}{2}}^{n+\frac{1}{2}+},$$

$$[F^1]_{i+}^{n+\frac{1}{2}+} \approx [F^1]_{i+\frac{1}{2}}^{n+\frac{1}{2}+} - \frac{1}{2}\Delta x^1 [X]_{i+\frac{1}{2}}^{n+\frac{1}{2}+},$$

$$[F^1]_{i-\frac{1}{2}}^{n+\frac{1}{2}+} \equiv \frac{1}{2}\left([F^1]_{i-\frac{1}{2}-}^{n+\frac{1}{2}-} + [F^1]_{i-\frac{1}{2}+}^{n+\frac{1}{2}-}\right)$$
$$- \frac{1}{2}\frac{\Delta x^0}{\Delta x^1}[F_u(F_u^0)^{-1}]_{i-\frac{1}{2}}^{n+\frac{1}{2}-}\left([F^1]_{i-\frac{1}{2}+}^{n+\frac{1}{2}-} - [F^1]_{i-\frac{1}{2}-}^{n+\frac{1}{2}-}\right)$$
$$\approx \frac{1}{2}\left([F^1]_{i-1}^{n+\frac{1}{2}-} + [F^1]_i^{n+\frac{1}{2}-}\right)$$
$$+ \frac{1}{2}\Delta x^0 \left[F_u^1(F_u^0)^{-1}\left(X - F_{,1}^1\right)\right]_{i-\frac{1}{2}}^{n+\frac{1}{2}-}$$
$$\approx \frac{1}{2}\left\{[F^1]_{i-1}^n + \frac{1}{2}\Delta x^0 \left[F_g^1 g_{,0} + F_u^1(F_u^0)^{-1}\left(T - F_g^0 g_{,0}\right)\right]_{i-1}^n\right.$$
$$\left. + [F^1]_i^n + \frac{1}{2}\Delta x^0 \left[F_g^1 g_{,0} + F_u^1(F_u^0)^{-1}\left(T - F_g^0 g_{,0}\right)\right]_i^n\right\}$$
$$+ \frac{1}{2}\Delta x^0 \left[F_u^1(F_u^0)^{-1}\left(X - F_{,1}^1\right)\right]_{i-\frac{1}{2}}^{n+\frac{1}{2}-},$$

$$[F^1]_{i+\frac{1}{2}}^{n+\frac{1}{2}+} \equiv \frac{1}{2}\left([F^1]_{i+\frac{1}{2}-}^{n+\frac{1}{2}-} + [F^1]_{i+\frac{1}{2}+}^{n+\frac{1}{2}-}\right)$$
$$- \frac{1}{2}\frac{\Delta x^0}{\Delta x^1}[F_u^1(F_u^0)^{-1}]_{i+\frac{1}{2}}^{n+\frac{1}{2}-}\left([F^1]_{i+\frac{1}{2}+}^{n+\frac{1}{2}-} - [F^1]_{i+\frac{1}{2}-}^{n+\frac{1}{2}-}\right)$$
$$\approx \frac{1}{2}\left([F^1]_i^{n+\frac{1}{2}-} + [F^1]_{i+1}^{n+\frac{1}{2}-}\right)$$
$$+ \frac{1}{2}\Delta x^0 \left[F_u^1(F_u^0)^{-1}\left(X - F_{,1}^1\right)\right]_{i+\frac{1}{2}}^{n+\frac{1}{2}-}$$
$$\approx \frac{1}{2}\left\{[F^1]_i^n + \frac{1}{2}\Delta x^0 \left[F_g^1 g_{,0} + F_u^1(F_u^0)^{-1}\left(T - F_g^0 g_{,0}\right)\right]_i^n\right.$$
$$\left. + [F^1]_{i+1}^n + \frac{1}{2}\Delta x^0 \left[F_g^1 g_{,0} + F_u^1(F_u^0)^{-1}\left(T - F_g^0 g_{,0}\right)\right]_{i+1}^n\right\}$$
$$+ \frac{1}{2}\Delta x^0 \left[F_u^1(F_u^0)^{-1}\left(X - F_{,1}^1\right)\right]_{i+\frac{1}{2}}^{n+\frac{1}{2}-}. \quad (D.1)$$

Likewise one has

$$[X]_{i-\frac{1}{2}}^{n+\frac{1}{2}+} \approx \frac{1}{2}\left([X]_{i-\frac{1}{2}-}^{n+\frac{1}{2}-} + [X]_{i-\frac{1}{2}+}^{n+\frac{1}{2}-}\right)$$
$$- \frac{1}{2}\frac{\Delta x^0}{\Delta x^1}[X_u(F_u^0)^{-1}]_{i-\frac{1}{2}}^{n+\frac{1}{2}-}\left([F^1]_{i-\frac{1}{2}+}^{n+\frac{1}{2}-} - [F^1]_{i-\frac{1}{2}-}^{n+\frac{1}{2}-}\right)$$
$$\approx \frac{1}{2}\left([X]_{i-1}^{n+\frac{1}{2}-} + [X]_i^{n+\frac{1}{2}-}\right)$$
$$+ \frac{1}{2}\Delta x^0 \left[X_u(F_u^0)^{-1}\left(X - F_{,1}^1\right)\right]_{i-\frac{1}{2}}^{n+\frac{1}{2}-}$$
$$\approx \frac{1}{2}\left\{[X]_{i-1}^n + \frac{1}{2}\Delta x^0 \left[X_g g_{,0} + X_u(F_u^0)^{-1}\left(T - F_g^0 g_{,0}\right)\right]_{i-1}^n\right.$$
$$\left. + [X]_i^n + \frac{1}{2}\Delta x^0 \left[X_g g_{,0} + X_u(F_u^0)^{-1}\left(T - F_g^0 g_{,0}\right)\right]_i^n\right\}$$
$$+ \frac{1}{2}\Delta x^0 \left[X_u(F_u^0)^{-1}\left(X - F_{,1}^1\right)\right]_{i-\frac{1}{2}}^{n+\frac{1}{2}-},$$

$$[X]_{i+\frac{1}{2}}^{n+\frac{1}{2}+} \approx \frac{1}{2}\left([X]_{i+\frac{1}{2}-}^{n+\frac{1}{2}-} + [X]_{i+\frac{1}{2}+}^{n+\frac{1}{2}-}\right)$$
$$- \frac{1}{2}\frac{\Delta x^0}{\Delta x^1}[X_u(F_u^0)^{-1}]_{i+\frac{1}{2}}^{n+\frac{1}{2}-}\left([F^1]_{i+\frac{1}{2}+}^{n+\frac{1}{2}-} - [F^1]_{i+\frac{1}{2}-}^{n+\frac{1}{2}-}\right)$$
$$\approx \frac{1}{2}\left([X]_i^{n+\frac{1}{2}-} + [X]_{i+1}^{n+\frac{1}{2}-}\right)$$
$$+ \frac{1}{2}\Delta x^0 \left[X_u(F_u^0)^{-1}\left(X - F_{,1}^1\right)\right]_{i+\frac{1}{2}}^{n+\frac{1}{2}-}$$
$$\approx \frac{1}{2}\left\{[X]_i^n + \frac{1}{2}\Delta x^0 \left[X_g g_{,0} + X_u(F_u^0)^{-1}\left(T - F_g^0 g_{,0}\right)\right]_i^n\right.$$
$$\left. + [X]_{i+1}^n + \frac{1}{2}\Delta x^0 \left[X_g g_{,0} + X_u(F_u^0)^{-1}\left(T - F_g^0 g_{,0}\right)\right]_{i+1}^n\right\}$$
$$+ \frac{1}{2}\Delta x^0 \left[X_u(F_u^0)^{-1}\left(X - F_{,1}^1\right)\right]_{i+\frac{1}{2}}^{n+\frac{1}{2}-}. \quad (D.2)$$

After substitution of these expressions one finds

$$[F^0]_i^{n+\frac{1}{2}+} \approx [F^0]_i^{n+\frac{1}{2}-} + \Delta x^0 \left[\left(X - F_{,1}^1\right)\right]_i^n + \frac{1}{2}(\Delta x^0)^2 \left\{\right.$$
$$X_g g_{,0} + X_u(F_u^0)^{-1}\left(X + T - F_{,1}^1 - F_g^0 g_{,0}\right)$$
$$\left. - \left[F_g^1 g_{,0} + F_u^1(F_u^0)^{-1}\left(X + T - F_{,1}^1 - F_g^0 g_{,0}\right)\right]_{,1}\right\},$$

$$[T]_i^{n+\frac{1}{2}+} \approx [T]_i^{n+\frac{1}{2}-} + \Delta x^0 T_u(F_u^0)^{-1}\left(X - F_{,1}^1\right)$$
$$\approx [T]_i^n + \frac{1}{2}\Delta x^0 \left[T_g g_{,0} + T_u(F_u^0)^{-1}\left(T - F_g^0 g_{,0}\right)\right]$$
$$+ \Delta x^0 T_u(F_u^0)^{-1}\left(X - F_{,1}^1\right). \quad (D.3)$$

After substitution of these expressions in Eq. (D.1) one can conclude that the terms of first and second order in $\Delta x^0$ are indeed equal to their analytical counterparts

$$\Delta x^0 [S - F_{,1}^1]_i^n \quad (D.4)$$

and

$$\frac{1}{2}(\Delta x^0)^2 \left\{S_g g_{,0} + S_u(F_u^0)^{-1}\left(S - F_{,1}^1 - F_g^0 g_{,0}\right)\right.$$
$$\left. - \left[F_g^1 g_{,0} + F_u^1(F_u^0)^{-1}\left(S - F_{,1}^1 - F_g^0 g_{,0}\right)\right]_{,1}\right\} \quad (D.5)$$

respectively.

**Appendix E: Characteristic form of the equations**

In this appendix it is shown that the present method applied in the limit of two (almost) identical interface states makes use of the exact characteristic form of the equations of general relativistic fluid dynamics projected onto one spatial dimension.

For two identical states the averaging in Eq. (10.7) becomes trivial:

$$<K> \to (\sqrt{-g}\rho h)^{\frac{1}{2}},$$
$$v^\alpha \to u^\alpha,$$
$$v^4 \to \frac{p}{\rho h},$$
$$s^2 \to \Gamma \frac{p}{\rho h}. \quad (E.1)$$

Substitution in the other expressions of Sect. 10 is now straightforward.

The hardest to check is the expression for the sound speed in arbitrary coordinates $x^\alpha$, given that in Lorentz-Minkowski coordinates $X^\alpha$ which are comoving with the instantaneous systematic fluid velocity the sound speed is $s$. Because the fluid is at rest in this frame,

$$u^\alpha = \frac{dx^\alpha}{d\tau} = \frac{\partial x^\alpha}{\partial X^\beta}\frac{dX^\beta}{d\tau} = \frac{\partial x^\alpha}{\partial X^0}. \quad (E.2)$$

Thus in terms of the systematic velocity $u^\alpha$

$$g^{\alpha\beta} + u^\alpha u^\beta = \eta^{ij}\frac{\partial x^\alpha}{\partial X^i}\frac{\partial x^\beta}{\partial X^j}. \quad (E.3)$$

Having established these relations, maximize the sound speed in the 1-direction over the spatial components of the rest frame sound speed,

$$\lambda = \frac{s^1}{s^0} = \frac{\frac{\partial x^1}{\partial X^\alpha}S^\alpha}{\frac{\partial x^0}{\partial X^\alpha}S^\alpha}, \quad (E.4)$$

under the conditions that the rest frame sound speed is $s$:

$$S^0 = \frac{1}{\sqrt{1-s^2}}, \qquad \eta_{ij}S^iS^j = \frac{s^2}{1-s^2}. \quad (E.5)$$

Using a Lagrange multiplier $\mu$ one obtains the equations for the spatial components of the rest frame sound speed

$$\frac{\partial x^1}{\partial X^j} - \lambda\frac{\partial x^0}{\partial X^j} = \mu S^j, \qquad j=1,2,3. \quad (E.6)$$

These relations can be substituted into the restrictions Eq. (E.5) to yield, after replacing the products of partial derivatives by the metric elements:

$$\mu^2\frac{s^2}{1-s^2} = (g^{00} + u^0u^0)\lambda^2 - (g^{01} + u^0u^1)2\lambda + (g^{11} + u^1u^1). \quad (E.7)$$

With this relation for the Lagrange multiplier applied to the spatial components of the rest frame sound speed, one obtains

$$(1-s^2)(u^0u^0\lambda^2 - 2u^0u^1\lambda + u^1u^1) = s^2(g^{00}\lambda^2 - 2g^{01}\lambda + g^{11}). \quad (E.8)$$

The roots of this equation are exactly equal to the roots $\lambda^\pm$ of Eq. (10.11) in the limit of small differences.

34  Frits Eulderink, Garrelt Mellema: General relativistic hydrodynamics with a Roe solver